%% file: main.tex
\documentclass[
    amsmath,%
	amsfonts,%
	amssymb,%
    pra,%
    aps,%
    groupedaddress,%
    superscriptaddress,%
    twocolumn,%
    toc=flat,%
    longbibliography,%
]{revtex4-2}

\usepackage[utf8]{inputenc}
\usepackage{xcolor}
\usepackage{bbold}
\usepackage{graphicx}
\usepackage{array,multirow}
\usepackage{boldline}
\usepackage{tabularx}
\usepackage{tikz}
\usepackage{ctable}
\usepackage{epsfig}
\usepackage[english, status=final]{fixme}
\fxsetup{marginface=\tiny}
\fxusetheme{color}

\usepackage[
    colorlinks=true,%
    linkcolor=blue,%
    urlcolor=blue,%
    citecolor=blue,%
]{hyperref}

\let\ldash\l
\renewcommand{\l}{\left(}

\newcommand{\U}{\hat{U}}
\newcommand{\B}{\hat{B}}
\newcommand{\G}{\hat{G}}
\newcommand{\Gt}{\tilde{G}}
\newcommand{\Ud}{\hat{U}^\dagger}
\newcommand{\Q}{\hat{Q}}

\newcommand{\bra}[1]{\langle#1|}
\newcommand{\ket}[1]{|#1\rangle}

\renewcommand{\ij}{{\langle i, j \rangle}}
\renewcommand{\H}{\hat{\mathcal{H}}}

\renewcommand{\c}{\hat{c}}
\renewcommand{\d}{\hat{d}}
\renewcommand{\a}{\hat{a}}
\newcommand{\dd}{\hat{d}^\dagger}
\newcommand{\cd}{\hat{c}^\dagger}
\newcommand{\ad}{\hat{a}^\dagger}
\newcommand{\bd}{\hat{b}^\dagger}
\renewcommand{\b}{\hat{b}}

\newcommand{\n}{\hat{n}}

\newcommand{\tr}{\text{tr}}

\newcommand{\taux}[2]{\hat{\tau}^x_{\langle{#1},{#2}\rangle}}
\newcommand{\tauz}[2]{\hat{\tau}^z_{\langle{#1},{#2}\rangle}}

\newcommand{\lrangle}[2]{\langle{#1},{#2}\rangle}
\newcommand{\adn}{\hat{a}^{\dagger (n)}}
\newcommand{\an}{\hat{a}^{(n)}}

\newcommand{\Ztwo}{$\mathbb{Z}_2$}

\newcommand*\curveminus{%
  \mathbin{\rotatebox[origin=c]{0}{$\curvearrowleft$}}}

\newcommand*\curveplus{%
  \mathbin{\rotatebox[origin=c]{180}{$\curvearrowright$}}}

\newdimen\satlevel
\newdimen\satdiameter
\newcommand{\satisfaction}[2][]{%
    \satdiameter=1.4ex\relax
    \ifcase#2\relax
        \satlevel=0pt\relax
    \or
        \satlevel=0.125\satdiameter
    \or
        \satlevel=0.25\satdiameter
    \or
        \satlevel=0.375\satdiameter
    \or
        \satlevel=0.5\satdiameter
    \fi
    \tikz[baseline=-0.4\satdiameter]{%
        \draw[#1] (0,0) circle (0.5\satdiameter);
        \fill[#1] (0,0) circle (\satlevel);
    }%
}
\usepackage{ragged2e}

\usepackage{bm}	
\renewcommand{\vec}[1]{\bm{#1}}

\begin{document}
\title{\Ztwo{} lattice gauge theories and Kitaev's toric code:\\
A scheme for analog quantum simulation}

\author{Lukas Homeier}
\thanks{These authors contributed equally.}
\affiliation{Department of Physics and Arnold Sommerfeld Center for Theoretical Physics (ASC), Ludwig-Maximilians-Universit\"at M\"unchen, Theresienstr.\ 37, D-80333 M\"unchen, Germany}
\affiliation{Munich Center for Quantum Science and Technology (MCQST), Schellingstr.\ 4, D-80799 M\"unchen, Germany}

\author{Christian Schweizer}
\thanks{These authors contributed equally.}
\affiliation{Fakult\"at f\"ur Physik, Ludwig-Maximilians-Universit\"at M\"unchen, Schellingstr.\ 4, D-80799 M\"unchen, Germany}
\affiliation{Munich Center for Quantum Science and Technology (MCQST), Schellingstr.\ 4, D-80799 M\"unchen, Germany}

\author{Monika Aidelsburger}
\affiliation{Fakult\"at f\"ur Physik, Ludwig-Maximilians-Universit\"at M\"unchen, Schellingstr.\ 4, D-80799 M\"unchen, Germany}
\affiliation{Munich Center for Quantum Science and Technology (MCQST), Schellingstr.\ 4, D-80799 M\"unchen, Germany}

\author{Arkady Fedorov}
\affiliation{ARC Centre of Excellence for Engineered Quantum Systems, Queensland 4072, Australia}
\affiliation{School of Mathematics and Physics, University of Queensland, St Lucia, Queensland 4072, Australia}

\author{Fabian Grusdt}
\affiliation{Department of Physics and Arnold Sommerfeld Center for Theoretical Physics (ASC), Ludwig-Maximilians-Universit\"at M\"unchen, Theresienstr.\ 37, D-80333 M\"unchen, Germany}
\affiliation{Munich Center for Quantum Science and Technology (MCQST), Schellingstr.\ 4, D-80799 M\"unchen, Germany}

\date{\today}

\begin{abstract}
Kitaev's toric code is an exactly solvable model with \Ztwo{}-topological order,
which has potential applications in quantum computation and error correction.
However, a direct experimental realization remains an open challenge. 
Here, we propose a building block for \Ztwo{} lattice gauge theories 
coupled to dynamical matter 
and demonstrate how it allows for an implementation 
of the toric-code ground state and its topological excitations. 
This is achieved by introducing separate matter excitations 
on individual plaquettes, 
whose motion induce the required plaquette terms. 
The proposed building block is realized in the second-order coupling regime 
and is well suited for implementations with superconducting qubits.
Furthermore, we propose a pathway to prepare topologically non-trivial initial states
during which a large gap on the order of the underlying coupling strength is present. 
This is verified by both analytical arguments and numerical studies.
Moreover, we outline experimental signatures of the ground-state wavefunction 
and introduce a minimal braiding protocol.
Detecting a $\pi$-phase shift between Ramsey fringes in this protocol
reveals the anyonic excitations of the toric-code Hamiltonian 
in a system with only three triangular plaquettes. 
Our work paves the way for realizing non-Abelian anyons in analog quantum simulators.
\end{abstract}

\maketitle

\section{Introduction}
Quickly after the integer quantum Hall effect had been theoretically understood, the richness of the unexpectedly discovered fractional quantum Hall effect has left no doubt that the addition of strong interactions can lead to even more remarkable topological phenomena \cite{Wen1991, Wen1999, Ichinose2014, Rachel2018}. These include topological ground-state degeneracies and anyonic excitations with non-Abelian braiding statistics. Over the past decades topological phases of matter have been extensively studied from a theoretical perspective~\cite{Thouless1982, Haldane1988, Hassan2010, Qi2011} and it has become a key challenge to directly observe and study these exotic states of matter in experiments~\cite{Ren2016}. Depending on the experimental platform, different obstacles have to be overcome. While the more traditional quantum Hall settings allow for an easy preparation of the required low-temperature states, it remains extremely challenging to exert fully coherent control over their topological excitations~\cite{Bonderson2006PRL96,Nakamura2020}. On the other hand, various analog quantum simulators, e.g.\ ultracold atoms, ions and superconducting qubits, have already demonstrated excellent coherent control over their individual constituents \cite{Gross2017, Blatt2012, Romero2016}. State preparation \cite{Popp2004,Hamma2008,Kapit2014,Grusdt2014,Barkeshli2015,Ni2016,Motruk2017,Wang2020} and the implementation of e.g.\ $N$-body interactions \cite{Zohar2013_2, Dai2017,Weimer2010,Hafezi2014,Glaetzle2014,Bohrdt2020}, however, remain challenging tasks. 

The concept of topological order is closely related to emergent gauge degrees-of-freedom. For example, the robust topological ground-state degeneracy on a torus can be understood as a result of non-local gauge excitations, which ultimately represent a non-trivial pattern of entanglement in the ground state. This close connection is most clearly demonstrated in Kitaev's toric code~\cite{Kitaev2003}, which represents an exactly solvable \Ztwo{} lattice gauge theory (LGT) \cite{Wegner1971}. Its ground state on a torus is $2^2=$ four-fold degenerate and has anyonic excitations with non-Abelian braiding statistics, which can be used for storing and processing quantum information \cite{Nayak2008}. These properties are universal and hold on an arbitrary 2D lattice. However, an experimental exploration of these fundamental concepts remains an open challenge.

Here, we propose a realistic scheme for an analog quantum simulation of Kitaev's toric code. Digital approaches have been described in~\cite{Song2018,Park2016,Zohar2017, Zohar2017_2,Bender2018} and analog ones in~\cite{Herdman2010, Zohar2013_2, Sameti2017, Verresen2020, Samajdare2021}. Our approach, however, is based on a general and scalable building block for a \Ztwo{} LGT coupled to matter, which relies on a second-order coupling scheme of harmonic and anharmonic oscillators. Hence, the scheme is well suited for an implementation with existing superconducting qubit technology. 

The remainder of this article is devoted to a detailed analysis of the proposed toric-code Hamiltonian on the triangular lattice. We first address the problem how the topologically non-trivial ground state can be prepared in a realistic setting of coupled superconducting qubits. We propose a growing scheme \cite{Grusdt2014}, which allows to adiabatically drive the system from a trivial product state into a topological phase, maintaining a large energy gap throughout. We next explain analytically how this scheme works and supplement our findings with numerical simulations. Then we discuss possible experimental signatures for detecting the topological phase, which are inspired by methods originally introduced in the context of quantum gas microscopy \cite{Kuhr2016}. Finally, we report a minimal braiding protocol of distinguishable elementary excitations in the system, which allows for a direct measurement of the non-trivial braiding phase.

The continuing development of superconducting qubit technology \cite{Krantz2019, Kjaergaard2020}
moves this platform towards the focus of quantum simulation applications. 
Superconducting qubit arrays have already been used to study 
the interplay of interactions and synthetic gauge fields \cite{Roushan2017,Vepsalainen2019,Vepsalainen2020}, 
many-body localization and the associated logarithmic entanglement growth \cite{Chiaro2019,Guo2020} 
as well as dissipatively stabilized Mott insulators \cite{Ma2019}, to name a few. 
Owing to the similarities of the various quantum simulation platforms, 
many of the results achieved, e.g.\ with ultracold atoms in optical lattices, have direct implications 
and can be carried over to the superconducting qubit platform. 
Indeed, in the past decade significant progress has been made 
engineering artificial gauge fields for neutral particles and photonics~\cite{Aidelsburger2018,Ozawa2019},
and combining them with strong interactions~\cite{Gemelke2010,Tai2017,Clark2020}
-- thus paving the way towards studies of strongly correlated topological states of matter~\cite{Anderson2016}.

\begin{figure}
\centering
\epsfig{file=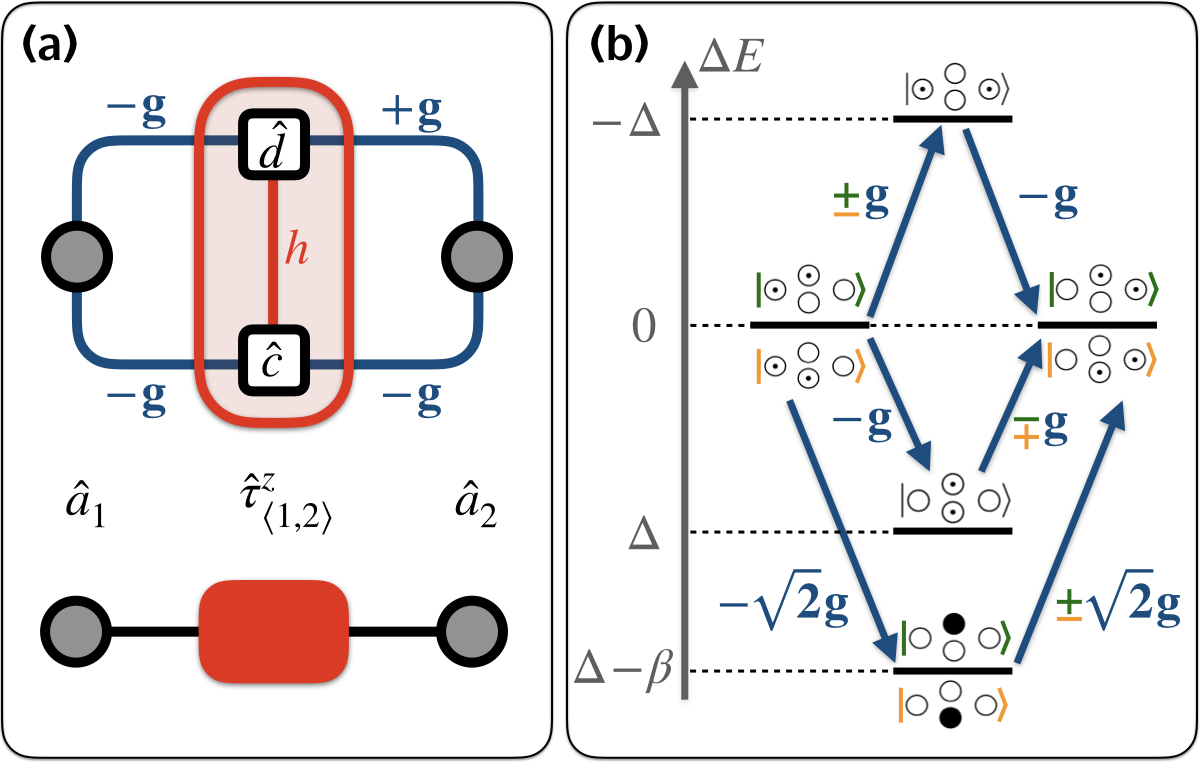, width=0.47\textwidth}
\caption{Building block in second-order coupling regime. (a) The two lattice sites (gray circles) 
are connected by a link with a \Ztwo{} gauge degree-of-freedom $\tauz{1}{2}$ (red box).
The lattice sites $\a_1$ and $\a_2$ are realized by harmonic resonators 
and $\tauz{1}{2}$ by two additional anharmonic oscillators $\c$ and $\d$ with anharmonicity~$\beta$, 
which are detuned to lower energy by~$\Delta<0$.
They are connected by couplings~$g$, one of which has opposite sign.
The sites $\c$ and $\d$ share a single excitation,
which realises the gauge degree-of-freedom $\tauz{1}{2} = \cd\c-\dd\d$.
(b) Second-order tunneling of a matter excitation from $\a_1$ to $\a_2$ 
can be achieved via three intermediate states.
The states are symbolized like $\ket{{\scriptstyle a_1}_{c}^{d}{\scriptstyle a_2}}$
with~$\satisfaction{0}$ zero, $\satisfaction{1}$ one,
and~$\satisfaction{4}$ two excitations on the respective oscillator;
orange and green kets mark $\tau^z=\pm 1$.
Note that depending on the gauge field,
one sign is reversed per second-order coupling process
leading to an effective coupling $(-t\,\ad_1 \tauz{1}{2} \a_2 + \mathrm{H.c.})$.
} 
\label{fig:coupling_scheme}
\end{figure}

In this article, we further establish superconducting qubit arrays 
as a promising platform for realizing \Ztwo{} LGTs coupled to dynamical matter fields \cite{Fradkin1979}. 
We show explicitly how qubits in a triangular lattice can be coupled in an elegant way 
to obtain local symmetries, 
which is characterized by almost perfectly conserved local constraints or \Ztwo{} Gauss' laws. 
Indeed, the elementary \Ztwo{} building block we propose for superconducting qubits, see Fig.~\ref{fig:coupling_scheme}, 
resembles the building block proposed \cite{Barbiero2019} 
and realized \cite{Schweizer2019} earlier with ultracold atoms in optical lattices, 
see also \cite{Goerg2019} and a proposal with two-species fermionic atoms~\cite{Zheng2020}. 
In contrast to those ultracold atom approaches, 
the scheme proposed here does not require Floquet engineering 
and therefore does not suffer from the notorious heating problem. 

This article is organized as follows: 
In Sec.~\ref{secImplementation} we introduce the elementary \Ztwo{} building block. 
We show in Sec.~\ref{secPlaquette} how building blocks can be combined 
and three-body plaquette interactions can be realized in a triangular lattice. 
Sec.~\ref{secToricCode} is devoted to the adiabatic preparation scheme, 
realistic experimental signatures and a discussion of the minimal braiding scheme. 
We close with a summary and outlook in Sec.~\ref{secSummary}.

\section{Implementation of gauge--matter coupling}
\label{secImplementation}
A \Ztwo{} lattice gauge theory coupled to matter 
is characterized by a \Ztwo{} gauge degree-of-freedom on every lattice link. 
When a matter excitation moves across such a link 
it picks up a $0$ or $\pi$-phase depending on the traversed link’s gauge field. 
The associated matter--gauge coupling Hamiltonian is: 
\begin{align}
    \H_{\mathbb{Z}_2} = -t \,\left( \ad_i \tauz{i}{j} \a_j +\mathrm{H.c.} \right),
    \label{eq:building_block}
\end{align}
where $t$ is the coupling strength, 
$\tauz{i}{j}$ is the gauge field on the link $\lrangle{i}{j}$ between site~$i$ and~$j$, 
and $\ad$ ($\a$) are the matter creation (annihilation) operators. 
Moreover, the motion of the matter particle
also changes the traversed link's gauge degree-of-freedom
according to the local Gauss' laws $[\G_i, \H_{\mathbb{Z}_2}] = 0$,
where $\G_i = \Q_i\prod_{j:\langle i, j \rangle} \taux{i}{j}$ is the local symmetry generator,
$\Q_i = (-1)^{\ad_i \a_i}$ is the \Ztwo{} charge,
and $\taux{i}{j}$ is the \Ztwo{} electric field.
For the presentation of the proposed experimental scheme, 
we restrict the description to a single building block. 
Extended models can be generated 
by connecting multiple building blocks together (see Appendix~\ref{supp:2nd_order}). 

The building block consists of two lattice sites connected by a link. 
Each lattice site is realized by a harmonic resonator 
expressed by $\a_1$ and $\a_2$, 
whose excitations define the matter excitations. 
The two sites are connected via two paths. 
On each path is an additional anharmonic oscillator
connected with~$|g|$ and energetically detuned  
to lower energy $\Delta<0$ and $|\Delta| \gg |g|$ 
with respect to the lattice-sites' resonator (Fig.~\ref{fig:coupling_scheme}a).
The associated creation (annihilation) operators of these additional anharmonic oscillators 
are $\cd$ ($\c$) and $\dd$ ($\d$). 
Note that the coupling from $d$ to $a_2$ has opposite sign~\cite{Filipp2011}.
In conclusion, the lattice sites are connected along both paths
by second-order processes of strength $|2g^2/\Delta E|$,
where $\Delta E$ is the energy difference of the initial or final to the virtual state.
The signs of these processes therefore depend on the sign of both 
$\Delta E$ and the involved tunneling events~$g$.

The characteristic \Ztwo{} gauge degree-of-freedom is realized 
by one excitation shared between $c$ and $d$
and the gauge field is defined as $\tauz{1}{2} = \cd\c-\dd\d$.
Hence, the electric field term is $\taux{1}{2} = \cd\d + \dd\c$ 
and can be implemented by a tunable coupling~$h$ between the anharmonic oscillators.
This tunable coupling can be realized by connecting the anharmonic oscillators $c$ and $d$
via a coupler circuit e.g.\ another anharmonic oscillator,
whose frequency can be externally controlled.
In this setting the effective coupling strength~$h$ is proportional to the sum of the inverse energy difference between the virtual and the initial, and the virtual and the final state
and covers a large enough tunability to reach both the trivial and topological regime.
Typically, during initialization a state within a single gauge sector is prepared,
which means each link is in either of the two eigenstates of $\taux{i}{j}$.
This product state of different links,
where each link is represented by a superposition of the excitation on $c$ and $d$,
can be prepared by first detuning $c$ with respect to $d$, 
placing one excitation on the energetically lower or higher site 
and subsequently making them resonant with an adiabatic parameter change, hence realizing $\tau^x_{\lrangle{i}{j}} = \pm 1$, respectively.
Note, the building block relies on coherent dynamics of this gauge degree-of-freedom
and therefore the lifetime of the excitation needs to be much longer than the experiment time.

The anharmonicity~$\beta$ of the oscillators $c$ and $d$
lead to an interaction between the matter and the gauge-field excitation.
For each eigenstate of $\tauz{1}{2}$, 
the building block has three virtual states (Fig.~\ref{fig:coupling_scheme}b).
The effective second-order coupling between the lattice site $a_1$ and $a_2$ 
is given by the sum of the individual processes, which yields $t = 2g^2\beta/(\Delta^2+\Delta\beta)$.
Due to the high symmetry of the scheme and the single reversed sign, 
the signs of all individual processes are opposite for the two eigenvalues $\tau^z_{\lrangle{1}{2}} = \pm 1$.
Moreover, all $\tauz{1}{2}$-dependent dispersive energy shifts vanish~\cite{Jerger2016}
and no fine-tuning of the values $\Delta$, $\beta$, and $g$ is required 
to fulfill the gauge symmetry;
however, within each building block the values need to be equal (see Appendix~\ref{supp:2nd_order}).
In general, a weak matter-occupation dependence of the second-order parameters remains,
which conserves the gauge symmetry.
By selecting suitable parameter triples $(\Delta, \beta, g)$ these contributions can be removed,
which we verified by numerical time evolution of
an initially localized matter excitation on a triangle (see Appendix~\ref{supp:numerics_triangle}).
The scheme is applicable to a variety of platforms, e.g.\ circuit quantum electrodynamics,
as it statically interconnects harmonic and anharmonic oscillators.
In conclusion, the scheme constitutes a scalable building block for Hamiltonian~\eqref{eq:building_block} 
in second-order perturbation theory.

\section{Realization of plaquette terms and effective Hamiltonian}
\label{secPlaquette}
\begin{figure}
\centering
\epsfig{file=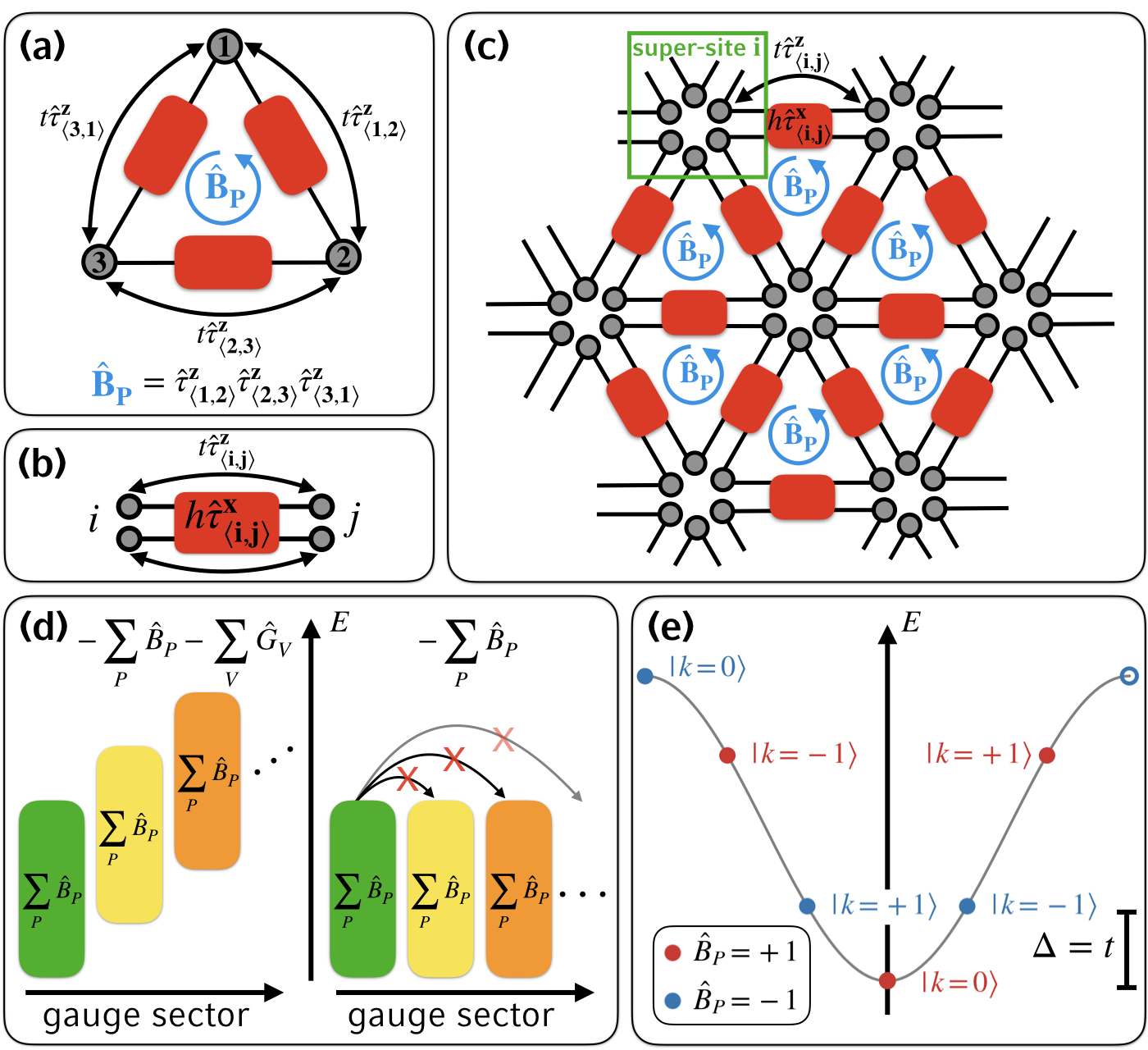, width=0.47\textwidth}
\caption{Microscopic model. (a) Matter sites (solid grey circles) are coupled to neighboring sites within a triangular plaquette $P$ by $t\,\tauz{i}{j}$. The motion of matter is restricted to a $1$D motion around $P$ to implement an effective plaquette operator $\B_P$. (b) To construct the full triangular lattice with restricted matter excitation, a double link with separated matter sites but a shared \Ztwo{} link variable is introduced. The latter can be supplemented by a \Ztwo{} electric term $h\,\taux{i}{j}$ of strength $h$. (c) On the full lattice, every triangular plaquette (a) is extended by the double link (b). The link variables $\hat{\vec{\tau}}_{\langle i,j\rangle}$ are located on the links between two super-sites $i$ and~$j$. After a gauge transformation the flux through a plaquette $P_n$ is given by the plaquette operator $\B_{P_n}$. (d) The toric code is a \Ztwo{} LGT and the physical Hilbert space can be decomposed into different gauge sectors (colored boxes). Vertex terms $\G_V = \prod_{\langle i,j \rangle \in V} \taux{i}{j}$ can be used to lift the degeneracy in the ground-state manifold between the sectors. Here, in contrast, we assume that coupling to other gauge sectors can be neglected on the relevant timescales. (e) The spectrum of Hamiltonian~(\ref{eq:single_triangle_hamiltonian}) with $h=0$ for the single triangle (a) depends on the plaquette eigenvalue $B_P=\pm1$. The two configurations are related by a phase shift $\pi$ along the cosine dispersion coming from the motional matter states labeled by~$k$. The ground state $B_P=+1 $ is gapped from the degenerate excited state $B_P = -1$ by $\Delta=t$.  } 
\label{fig:mic_model}
\end{figure}

In  the following, we will use the building block from Eq.~\eqref{eq:building_block} to construct a \Ztwo{} LGT on a triangular lattice and show that the ground state of this model resembles the topologically-ordered ground state of the toric-code Hamiltonian~\cite{Kitaev2003}. Specifically, we will show how the following Hamiltonian,
\begin{align}
    \H_\mathrm{tc} = -t\sum_n \B_{P_n},
    \label{eq:toric_code_plaquette}
\end{align}
can be effectively realized, where $\B_{P_n}=\prod_{\lrangle{i}{j}\in P_n} \tauz{i}{j}$ is the plaquette operator and $P_n$ labels plaquettes on the lattice. For a plaquette~$P$, we will show that the interaction $\B_P$ can be mediated by the motion of a single matter excitation confined to hop only around $P$ (Fig.~\ref{fig:mic_model}a).

To construct a full $2$D lattice such that matter excitations remain on their respective plaquette, we introduce a double-link element (Fig.~\ref{fig:mic_model}b) that plugs together individual plaquettes (Fig.~\ref{fig:mic_model}c). The resulting model has toric-code properties, which are revealed after matter and gauge fields are disentangled by an exact basis transformation~$\U$ introduced below. 
We will show that in this new basis, the Gauss' laws from the initial \Ztwo{} LGT building block yield new local symmetries that are identical to the toric-code vertex terms $\G_V = \prod_{\langle i,j \rangle \in V} \taux{i}{j}$. The ground state of the toric code then coincides with the ground state of the unique, appropriately chosen gauge sector in our proposed model as illustrated in Fig.~\ref{fig:mic_model}d. 
In the following, we assume that gauge-symmetry breaking couplings are small
and can be neglected on relevant timescales in experiments.

\subsection{Single plaquette}
We will first consider one triangular plaquette and show how the $\B_P$ operator arises. The Hamiltonian on a single plaquette is constructed from three \Ztwo{} gauge-matter building blocks~(\ref{eq:building_block}) with exactly one bosonic matter excitation (Fig.~\ref{fig:mic_model}a):
\begin{align}
\label{eq:single_triangle_hamiltonian}
\begin{split}
       \H_{\triangle} = &-t\sum_{\lrangle{i}{j}} \left( \ad_i \tauz{i}{j} \a_j + \mathrm{H.c.} \right) + h\sum_{\lrangle{i}{j}}\taux{i}{j}. 
\end{split}
\end{align}
The Hamiltonian has local, operator-valued hopping amplitudes $t\,\tauz{i}{j}$ 
and an additional term, which couples to the \Ztwo{} electric field with strength $h$. The model contains a \Ztwo{} lattice gauge structure with local symmetry generators $[\G_i,\mathcal{\hat{H}}]=0$, where $\G_i=(-1)^{\n_i}\prod_{j:\lrangle{i}{j}}\taux{i}{j}$. The number operator $\n_i$ counts the number of matter excitations on site $i$.

To derive the $\B_P$-dependent Hamiltonian~(\ref{eq:toric_code_plaquette}), we start with $\H_\triangle(h=0)$. The construction is inspired by the idea that the matter excitation acquires a phase $\hat{\Phi}$ when hopping around the plaquette, which is determined by the configuration of the \Ztwo{} link variables $\tauz{i}{j}$. The eigenstates of the Hamiltonian thus depend on the phases $\hat{\Phi}[\tauz{i}{j}]$ and effectively yield $\B_P$ terms in the energy.

As a first step, we introduce a basis transformation $\U$ that distributes the local hopping phases equally among the matter sites via an operator-valued phase shift, $\Ud \a_j \U = \a_j e^{i\hat{\vartheta}_j[\hat{\tau}^z]}$ with 
\begin{align}
   \label{eq:gauge_trafo_explicit}
   \hat{\vartheta}_j[\hat{\tau}^z] &=  \frac{\pi}{2}\left( \hat{\tau}^z_{\langle j,j+1 \rangle} - \hat{\tau}^z_{\langle j-1,j \rangle} \right),\\
   \U &= e^{i\sum_j \hat{\vartheta}_j[\hat{\tau}^z]\ad_j \a_j}.
\end{align}
In the following, quantities in the new basis are labeled by $\ket{\tilde{\psi}} = \U \ket{\psi}$. In the transformed basis and written in momentum representation $\a_{k_m} = 3^{-1/2} \sum_{j}e^{-ik_m R_j}\a_{j}$ of the matter sites, Hamiltonian~(\ref{eq:single_triangle_hamiltonian}) is given by
\begin{align}
    \Ud \H_\triangle |_{h=0} \U &= -2t\sum_{k_m = \frac{2\pi}{3}m}\cos(k_m + \hat{\Phi})\ad_{k_m} \a_{k_m} \label{eq:single_triangle_ft}\\
    \hat{\Phi} &= 
    \begin{cases}
        0,& \hat{B}_{P} = 1\\
        \pi, &  \hat{B}_{P} = -1,
    \end{cases} \label{eq:phase_plaquette}
\end{align}
where $k_m$ ($m=0,\pm 1$) labels discrete states in momentum space. Hamiltonian~(\ref{eq:single_triangle_ft}) is exactly solved by a product state $\ket{\tilde{\psi}} = \ket{\tilde{\psi}}_\mathrm{matter}\otimes\ket{\tilde{\psi}}_\mathrm{links}$ and the spectrum is given by the $\hat{\Phi}$-dependent dispersion as plotted in Fig.~\ref{fig:mic_model}e. The low-energy manifolds are gapped by $\Delta=t$ and directly implement the $\B_P$ dependency in the Hamiltonian. Note that the excited manifold $\B_P=-1$ has a two-fold degeneracy due to the motional freedom $k_m=\pm 2\pi/3$ of the matter excitation.

We emphasize that the triangular geometry plays a crucial role in the transformation $\U$. With only three matter sites, the phase can be distributed such that the total flux $\hat{\Phi} \propto \bigl( \B_P \bigr)^3 = \B_P$ introduces a large $\pi$ phase shift of the cosine dispersion. This leads to relatively large energy gaps, on the order of the underlying energy scale $t$, between discrete states with $B_P=\pm 1$. Such strong three-site interaction terms make the scheme appealing for experimental realization.

\subsection{Multiple plaquettes: triangular lattice}
As a next step, the full triangular lattice can be constructed by combining individual plaquettes. To avoid one plaquette influencing another, the matter excitations realizing individual $\B_P$ terms will be constrained to move around their respective plaquettes only. Since neighboring plaquette operators $\B_P$ share a \Ztwo{} gauge variable, we introduce the double-link building block shown in Fig.~\ref{fig:mic_model}b: It couples two independent matter fields $\a^{(1)}$ and $\a^{(2)}$, on opposite sides, to the same shared \Ztwo{} gauge field, $\H_{\mathbb{Z}_2}^{(2)} = -t \bigl( \a_i^{\dagger (2)} \tauz{i}{j} \a_j^{(2)} + \a_i^{\dagger (1)} \tauz{i}{j} \a_j^{(1)} + \mathrm{H.c.} \bigr)$, and can be supplemented by the \Ztwo{} electric field term $h\,\taux{i}{j}$.

The full $2$D triangular lattice can now be constructed as shown in Fig.~\ref{fig:mic_model}c. The cluster of matter sites that belong to each vertex on the triangular lattice will be called super-site. The Hamiltonian of the model is 
\begin{align}
\label{eq:mic_hamiltonian}
\begin{split}
       \H = &-t\sum_{n}\sum_{\lrangle{i}{j}\in P_n} \left(\adn_i \tauz{i}{j} \an_j + \mathrm{H.c.}\right) \\ 
       &+ h\sum_{\lrangle{i}{j}}\taux{i}{j}.
\end{split}
\end{align}
Note that hopping of the matter excitations $\an_i$ is constrained to a $1$D motion within a single plaquette $P_n$, and the number of matter excitations is restricted to exactly one boson per plaquette $P_n$: $\sum_{i \in P_n} \adn_i \an_i = 1$.

The local symmetry generators from the single triangle can be generalized to super-site operators $\G_i=(-1)^{\hat{N}_i}\prod_{j:\lrangle{i}{j}}\taux{i}{j}$. Here the super-site number operator $\hat{N}_i$ counts all matter excitations on the individual sites that belong to a given super-site~$i$ (Fig.~\ref{fig:mic_model}c) and we define $N^P_i$ as the number of plaquettes connected to super-site~$i$. Furthermore, the basis transformation $\U$ does not have to be extended but still acts on the individual sites Eq.~(\ref{eq:gauge_trafo_explicit}) instead of super-sites (Appendix~\ref{supp:eff_hamiltonian}). The transformed Hamiltonian~(\ref{eq:mic_hamiltonian}) and Gauss' laws $\Gt_i$ then result in:
\begin{align}
    \Ud \H |_{h=0} \U 
    &=\sum_{n}\sum_{k_m = \frac{2\pi}{3}m} \cos{ \left(k_m + \hat{\Phi}^{(n)}\right)} \a^{\dagger(n)}_{k_m}\a^{(n)}_{k_m}
    \label{eq:hamiltonian_U}
    \\
    \Ud \G_i \U &:= \Gt_i = (-1)^{N^P_i}\prod_{j:\lrangle{i}{j}}\taux{i}{j}.
    \label{eq:red_gauss_law}
\end{align}
Here, the phase shift $\hat{\Phi}^{(n)}$ depends on the plaquette operators $\B_{P_n}$ on individual plaquettes as in Eq.~(\ref{eq:phase_plaquette}). The transformed Gauss' law $\Gt_i = (-1)^{N^P_i} \Gt_{V_i}$ resembles vertex operators $\Gt_{V_i}$ of the toric code up to a fixed pre-factor $(-1)^{N^P_i}$. In the bulk $N^P_i=6$, but odd values of $N^P_i$ can arise at the edges of the system.

In the free system with $h=0$, we can thus conclude that the many-body ground state of Eq.~\eqref{eq:mic_hamiltonian} has (i)~$B_{P_n}=+1$ and $k_m=0$ for each plaquette $P_n$ and (ii)~an emergent \Ztwo{} Gauss' law, i.e.\ the Hilbert space fragments into distinct gauge sectors as shown in Fig.~\ref{fig:mic_model}d. The Gauss' law can be freely chosen to be $\Gt_i = \pm 1$ for all~$i$ by a proper state preparation sequence, see Sec.~\ref{subsecStatePrep}, and the many-body eigenfunctions can be disentangled into a product form
\begin{equation}
 \ket{\tilde{\psi}}= \ket{\tilde{\psi}}_{\mathrm{matter},P_1}\otimes\dotsb \otimes \ket{\tilde{\psi}}_{\mathrm{matter},P_n}\otimes\ket{\tilde{\psi}}_\mathrm{links}.   
\end{equation}

In the following, we will consider $\Gt_{V_i} = (-1)^{N^P_i}\Gt_i= +1$ as in Kitaev's work \cite{Kitaev2003}. In the ground state of this sector $\ket{\tilde{\psi}}_\mathrm{links} \equiv \ket{\Psi_{\rm tc}}$ is identical to the topologically-ordered toric-code ground state $\ket{\Psi_{\rm tc}}$.

\subsection{\Ztwo{} electric term}
\label{secZ2electric}
The basis transformation $\U$ does not commute with the \Ztwo{} electric field terms $\taux{i}{j}$ and thus these terms transform non-trivially. So far, the terms were neglected by setting $h=0$ in Hamiltonian~(\ref{eq:single_triangle_hamiltonian}) and~(\ref{eq:hamiltonian_U}). Here, we will show that the $\taux{i}{j}$ terms in the new basis do not couple between different gauge sectors $\Gt_i$ (Fig.~\ref{fig:mic_model}d) and therefore $h\,\taux{i}{j}$ is a useful tuning parameter for adiabatic ground-state preparation (Sec.~\ref{secToricCode}).

The transformed \Ztwo{} electric field term (Appendix~\ref{supp:eff_hamiltonian}) reads
\begin{align}
    \Ud \taux{i}{j} \U = (-1)^{\Delta \n}\taux{i}{j},
    \label{eq:taux_trafo}
\end{align}
where $\Delta \n$ counts the imbalance of matter excitations between the two sides of the link $\ij$ but only takes into account matter sites $\a_i^{(n)}$ which are directly attached to the link variable $\taux{i}{j}$. Thus, Eq.~(\ref{eq:taux_trafo}) is indeed gauge invariant and has a non-trivial dependence on \Ztwo{} charges $\hat{Q}_i = (-1)^{\hat{n}_i}$ in the system. In fact, the model~(\ref{eq:mic_hamiltonian}) cannot be solved by a simple product wavefunction ansatz for $h \neq 0$. Nevertheless, for the two limiting cases $h/t \gg 1$ and $h=0$, the ground state is known to be in the trivial and topological phase, respectively. We will show that by tuning the parameters $h$ and $t$ appropriately, the system can be adiabatically transformed between the phases while the system remains in the initially chosen gauge sector.

\section{Preparation and probes of the topological toric-code phase}
\label{secToricCode}
The \Ztwo{} LGT coupled to matter on the triangular lattice, Eq.~\eqref{eq:mic_hamiltonian}, has both a topological, for $|h| \ll |t|$, and a trivial phase, for $|h|\gg |t|$, see Sec.~\ref{secPlaquette}. While the topological phase is interesting to study experimentally but hard to access, the trivial phase is easy to prepare as its ground state is a product state. In the following section, we propose a growing scheme \cite{Grusdt2014} for Hamiltonian~(\ref{eq:mic_hamiltonian}), which adiabatically connects the two phases in order to prepare a topologically-ordered ground state. Furthermore, a realistic detection scheme for the toric-code ground state is presented, as well as a protocol to extract anyonic braiding statistics. The results are underlined using numerical exact diagonalisation (ED) studies.

\subsection{State preparation}
\label{subsecStatePrep}
\begin{figure*}
\centering
\epsfig{file=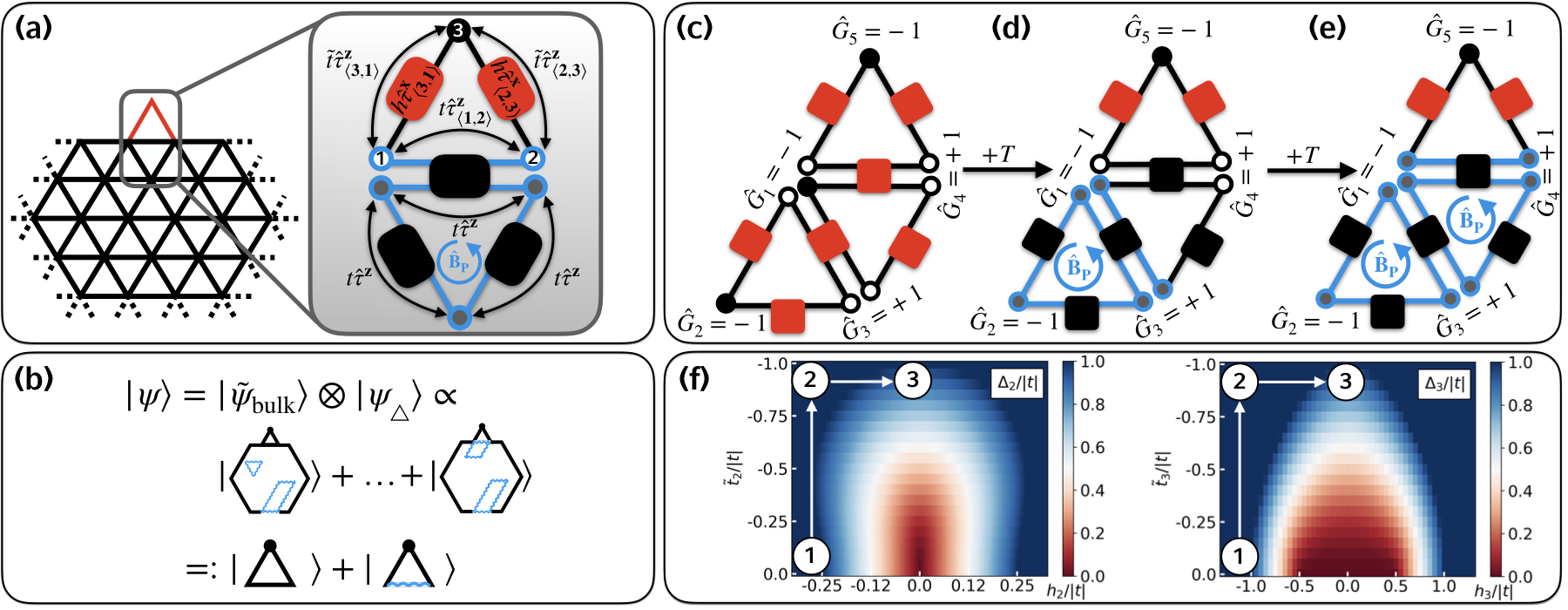, width=\textwidth}
\caption{State preparation by adiabatic growing scheme. (a) The bulk (black) is in the toric-code phase and the single plaquette (red) is initially in the trivial phase, i.e.\ link variables are in the $\tau^x=+1$ state stabilized via external, tunable coupling $h\,\taux{i}{j}$ (red boxes) and the matter excitation (black circle) is localized. Hopping $t\,\tauz{i}{j}$ is at full strength for the links connected to the bulk (blue lines) while hopping along the edge links $\tilde{t}\,\tauz{i}{j}$ can be tuned. During the growing step, the parameter $\tilde{t}$ ($h$) is increased (decreased) from zero to full strength (vice versa). (b) The terms described in (a) cannot couple between any eigenstates in the bulk. The Hamiltonian can be exactly expressed with reduced basis states. Here, the initial state of the growing scheme is shown in the reduced basis as an example. Straight (wiggly) lines denote the $\tau^x = +1$ ($\tau^x = -1$) state. (c) We illustrate the growing scheme for a small system using ED: The system is initialized in a trivial state. Black (empty) dots describe (the absence of) a localized matter excitation. The link variables are in the $\tau^x=+1$ state and the location of matter excitations is determined by $\G_i = (-1)^{\hat{N_i}} = (-1)^{N^P_i}$. This ensures growing into Kitaev's toric-code ground state. (d-e) In each growing step, one plaquette is added to the topological bulk.  (f) The plots show the many-body gap $\Delta_n/|t|$ for growing of the $n$-th plaquette versus the tunable parameters $\tilde{t}_n$ and $h_n$ used for adding the corresponding plaquette. The left plot corresponds to (d)$\rightarrow$(e) and the right plot shows the gap for growing the third plaquette. Drawn are suggested parameter paths with finite, constant gap $\Delta_n/|t|=1$.} 
\label{fig:growing_scheme}
\end{figure*}

We propose a growing scheme that starts in the trivial phase, i.e.\ all link variables $\taux{i}{j}$ are in the $\tau^x=+1$ eigenstate and the matter excitations are localized. By adiabatically turning on tunnelings $t$ -- plaquette after plaquette -- the system follows its ground state into the topologically-ordered toric-code state~\cite{Grusdt2014, Letscher2015}. We find that the scheme maintains a large gap $\Delta=t$ throughout the adiabatic evolution through parameter space. Such large gaps are a great advantage for experimental implementations since residual excitations are suppressed and the required time scales for state preparation scale polynomially with system size. We show this in a general way using analytical arguments and underline it with exact numerical studies, which demonstrate that state preparation with high fidelity is possible.

\emph{General procedure.--}
First, we discuss an individual growing step of the procedure. Initially, the system's bulk is in the toric-code phase and a single plaquette is in the trivial phase on the boundary (Fig.~\ref{fig:growing_scheme}a). With an adiabatic growing step the system is then transferred into the final state, in which the entire system is topologically ordered. This procedure can be repeated to grow systems of, in principle, arbitrary size. 

In the initial state, hoppings $t$ across links in the bulk are at full strength whereas hoppings $\tilde{t}=0$ across the two edge links of the new plaquette are switched off. On the two edge links, an external electric field term $h\,\taux{i}{j}$ stabilizes the link variables in the trivial phase, i.e.\ in the $\tau^x = +1$ state, and the matter excitation is pinned between those two links  (Fig.~\ref{fig:growing_scheme}a). 

The initial state is then the ground state of Hamiltonian~(\ref{eq:mic_hamiltonian}) with the described parameters. The growing step involves two consecutive steps: first, hopping $\tilde{t}$ is increased from $\tilde{t} = 0$ to $\tilde{t} = t$ and afterwards the external field $h$ is decreased to $h=0$. The final state is then the topologically-ordered ground state of Hamiltonian~(\ref{eq:hamiltonian_U}).

For the adiabatic growing scheme to work efficiently, the energy gap between the ground-state manifold and the excited state has to be large throughout the parameter path. In the following, we show this with analytical arguments. Since the Hamiltonian has no term that couples to excited bulk states, we reduce the basis states and solve the problem exactly in the reduced basis. Therefore, we decompose the states into $\ket{\psi} = \ket{\psi_{\tau^x=1}} + \ket{\psi_{\tau^x=-1}}$, where $\tau^x=\pm 1$ is the eigenvalue of the link variable connecting the bulk and the boundary triangle (Fig.~\ref{fig:growing_scheme}b). In this decomposition we still take into account all states of the boundary plaquette while the complete bulk Hilbert space can be reduced to its ground state without neglecting any couplings. A calculation shows that indeed an energy gap $\Delta/t = 1$ can be maintained throughout the growing step. For a detailed discussion see Appendix~\ref{supp:growing_scheme}.

With a sequence of growing steps an entire system can be prepared in the toric-code ground state. A fast growing procedure to prepare a large bulk with $N$ plaquettes could start with a minimal system -- a `crystal nucleus' -- around which hexagonal rings are grown simultaneously. For a given fixed fidelity per plaquette $F$ the required time per growing step is $t_F$. Our analysis above yields a short time scale $t_F \simeq 1/\Delta = 1/t$. The total time $T$ then scales as (see also Ref.~\cite{Hamma2008})
\begin{align}
    T \propto \frac{3}{2}t_F \sqrt{N}.
\end{align}
This polynomial scaling is much better compared to a generic exponential scaling for a system that is globally driven through the phase transition~\cite{Sameti2017}.

\begin{figure}
\centering
\epsfig{file=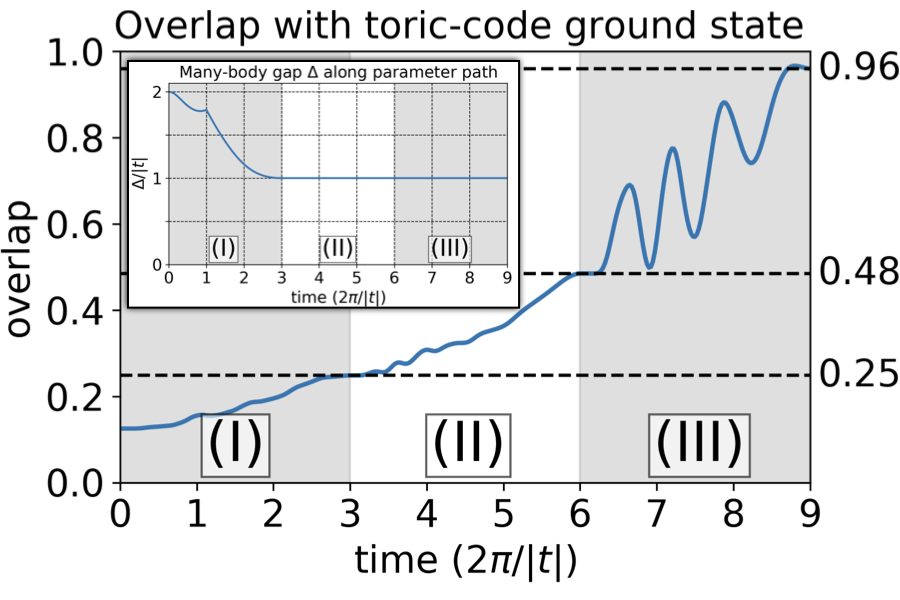, width=0.47\textwidth}
\caption{Numerical study of growing scheme with three plaquettes. The adiabatic time-evolution was calculated by ED. The system is initialized and then grown as explained in Fig.~\ref{fig:growing_scheme}c-e. Sections (I)-(III) indicate the growing of plaquette $P_1$-$P_3$. We plot the overlap $\tr_\mathrm{links}[\tr_\mathrm{matter}( \ket{\tilde{\psi}(t)}\bra{\tilde{\psi}(t)})\ket{\Psi_\mathrm{tc}}\bra{\Psi_\mathrm{tc}}]$, where $\ket{\tilde{\psi}(t)} := \U \ket{\psi(t)}$ is the time-evolved state in the new basis at time~$t$ and $\ket{\Psi_\mathrm{tc}}$ is the desired toric-code ground state. The overlap at $t_\mathrm{final}$ is larger than $96\%$. The plot in the inset shows the constantly large gap along the adiabatic sweep. During the preparation of plaquette $P_1$ the gap $\Delta/t > 1$ even exceeds~$t$.} 
\label{fig:time_evolution}
\end{figure}

\emph{Small system ED study.--}
In the following, we illustrate the growing scheme for a small system with three plaquettes, i.e.\ seven links and nine matter sites (Fig.~\ref{fig:growing_scheme}c-e).
We initialize the system with all link variables in the $\tau^x=+1$ state, which determine the unique positions of the matter excitations via the microscopic Gauss' laws $\G_i=(-1)^{\hat{N}_i} = (-1)^{N^P_i}$. The sign of the $\G_i$'s ensures that after the transformation $\U$ the vertex operators $\Gt_{V_i} = +1$ are positive (Eq.~\ref{eq:red_gauss_law}). Next, additional plaquettes can be adiabtically grown step by step.

We analyze the proposed growing scheme for the microscopic Hamiltonian~(\ref{eq:mic_hamiltonian}) using ED. First, the analytical calculations of the energy gap $\Delta$~(Appendix~\ref{supp:growing_scheme}) are verified for each growing step. Fig.~\ref{fig:growing_scheme}f shows the gaps $\Delta_2$ ($\Delta_3$) for growing the second (third) plaquette. The many-body spectrum has a constant gap $\Delta/|t|=1$ within connected areas in the parameter landscape $(\tilde{t},h)$. The suggested parameter path only contains points at which the gap is constant and open.

Secondly, the system is time-evolved in three consecutive growing steps for each plaquette. The parameter path is chosen as indicated by the white arrows in Fig.~\ref{fig:growing_scheme}f for each step and the parameters are ramped linearly in time. To extract the fidelity of our growing scheme, we calculate the overlap between the actual state and the desired toric-code ground state, $\tr_\mathrm{links}[\tr_\mathrm{matter}(\ket{\tilde{\psi}(t)}\bra{\tilde{\psi}(t)})\ket{\Psi_\mathrm{tc}}\bra{\Psi_\mathrm{tc}}]$. Here,~$\ket{\Psi_\mathrm{tc}}$ is the ground state of Hamiltonian~(\ref{eq:toric_code_plaquette}) in the gauge sector~$\G_{V_i}=+1$ on a system with three plaquettes. Fig.~\ref{fig:time_evolution} shows the results of the time-evolution by ED. The exponential growth of the overlap is in agreement with the growing occupation of the Hilbert space by a factor of two after each growing step. For the chosen duration of the adiabatic sweeps in our calculation, an overlap of over $96\%$ can be obtained for three plaquettes. The success of the adiabatic scheme relies on the finitness of the many-body gap along the parameter path that can be achieved by the stepwise growing scheme.

\subsection{Experimental signatures}
\begin{figure}
\centering
\epsfig{file=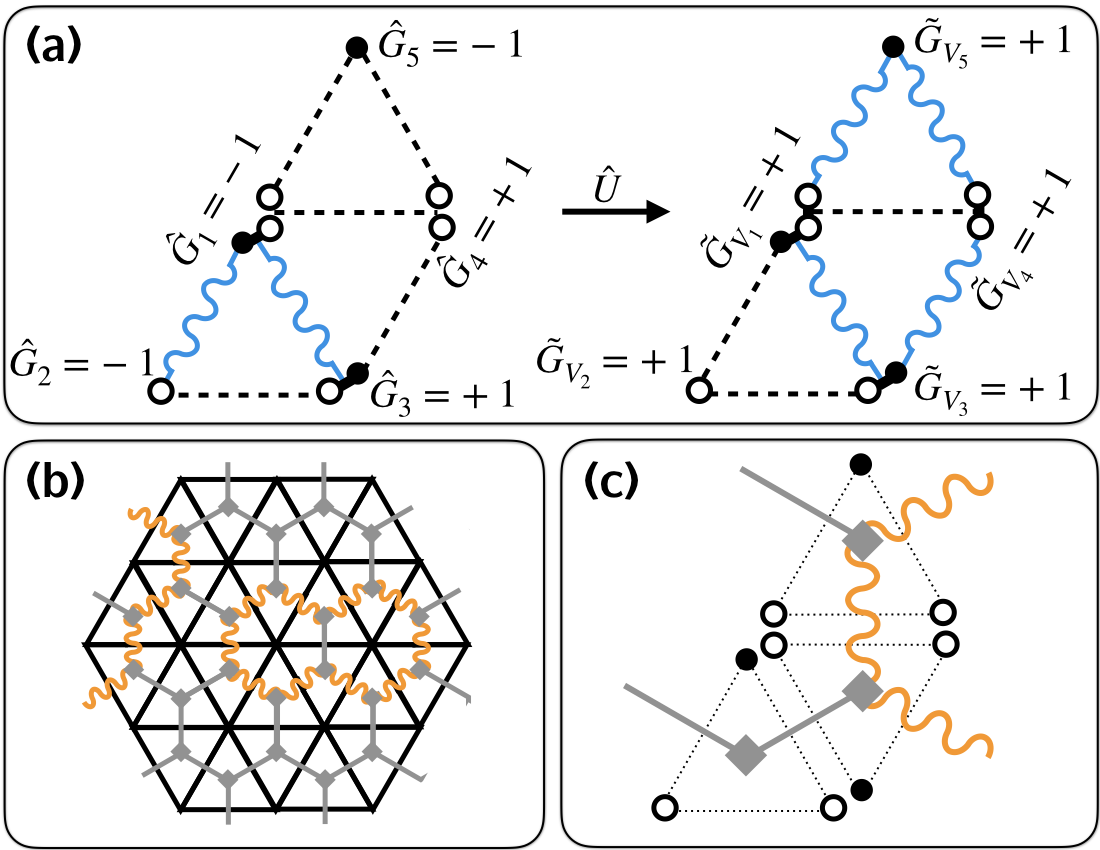, width=0.47\textwidth}
\caption{Snapshots. (a) A dotted black (wiggly blue) line is $\tau^x_{\langle i,j \rangle}=+1$ ($\tau^x_{\langle i,j \rangle}=-1$) in the string language. The left side of (a) shows a snapshot taken in the laboratory frame. The basis transformation $\U$ can flip strings depending on the configuration of matter excitations (right side). In the new basis, the vertex operators $\G_{V_i} = +1$ are manifested in closed loops of strings. (b-c) To measure the ground-state condition, $\B_{P_n}=+1$, snapshots have to be taken in the dual $\hat{\tau}^z$ basis. Lattice sites of the dual, honeycomb lattice are indicated by grey squares and the boundaries are open, i.e.\ the dual links do not end on dual lattice sites at the boundary. Here, a grey straight (wiggly orange) line is $\tau^z_{\langle i,j \rangle}=+1$ ($\tau^z_{\langle i,j \rangle}=-1$) in the string language. The system is in a state with all $\B_{P_n}=+1$ when for every measurement no string ends on a dual lattice site. } 
\label{fig:snapshot}
\end{figure}

The direct detection and verification of many-body quantum states often represents a very challenging task in experiments. Due to the nature of topological phases, there are no local order parameters to characterize the phase. With the development of analog quantum simulation platforms, projection measurements on the single-particle level have become possible and can be used to reconstruct the many-body wavefunction. In the following, we present possible experimental signatures to detect the toric-code ground-state wavefunction, for which we use the basis transformation $\U$ to reveal the hidden topological order in the microscopic system~\cite{Endres2011, Hilker2017}.

In \Ztwo{} LGTs, a string picture can be introduced by defining a string (no string) as a link variable in the $\tau^x = -1$ ($\tau^x = +1$) state. The toric-code ground state is then characterized by a superposition of states that only contain closed loop configurations of strings. In the \Ztwo{} LGT coupled to matter~(\ref{eq:mic_hamiltonian}), however, the Gauss' law $\G_i = +1$ allows strings to have an open end at super-site~$i$ in the presence of an odd number of matter excitations on super-site $i$, i.e.\ in the presence of a super-site charge $\hat{Q}_i$. Strings can be moved along the lattice by hopping of $\hat{Q}_i$, to which the strings are attached. Nevertheless, the toric-code ground state -- with its closed strings -- is revealed in our system after the basis transformation $\U$ (Eq.~\ref{eq:gauge_trafo_explicit}) has been applied as discussed in Sec.~\ref{secPlaquette}. The transformation $\U$ can flip the strings, where the flipping depends on the occupation of matter sites attached to $\taux{i}{j}$. By measuring the local matter excitations and string configurations in the laboratory basis, the transformation $\U$ can be fully evaluated (Eq.~\eqref{eq:taux_trafo}).

By evaluating snapshots taken in the laboratory frame, we determine whether the system is in the gauge sector ${\G_{V_i}=+1}$. Taking a snapshot in the laboratory frame automatically leads to string configurations that can have open ends (Fig.~\ref{fig:snapshot}a left). After evaluating the basis transformation $\U$, however, some strings are flipped yielding a new string configuration (Fig.~\ref{fig:snapshot}a right). In the new basis, the toric-code vertex operators $\G_{V_i}$ can be calculated to verify that the targeted gauge sector has been realised; for $\G_{V_i} = +1$ the closed string configurations can be seen (see also Appendix~\ref{supp:hidden_order}).

To completely determine the ground-state properties, the plaquette terms $\B_{P_n}$ need to be evaluated, which requires measurements in the $\hat{\tau}^z$ basis. We use the duality property of the toric-code model and proceed in a similar fashion as before. For this we identify plaquettes (vertices) of the triangular lattice with the vertices (plaquettes) of the dual, honeycomb lattice as defined in Fig.~\ref{fig:snapshot}b-c. In the dual basis, the ground-state wavefunction is then given by the closed string configurations on the honeycomb lattice measured in the $\hat{\tau}^z$ basis. Since the link variables $\Ud \tauz{i}{j} \U= \tauz{i}{j}$ do not change under the basis transformation, measurements can be directly performed in the laboratory basis. Since we started with closed boundaries on the triangular lattice, the dual lattice now has open boundaries and closed loops are defined by strings that do not end on dual lattice sites. Thus, strings can end on the open boundaries.

To conclude, repeated state preparations and projective measurements (snapshots) on the single sites as well as link variables can be used to reconstruct the state of the system with sufficient statistics. For the toric-code ground state, closed loops of strings measured in the $\hat{\tau}^x$ and $\hat{\tau}^z$ basis completely determine the $\G_V=+1$ and $\B_P=+1$ configurations.

\subsection{Topological excitations and minimal braiding scheme}
\label{secTopExcBraiding}
\begin{figure}
\centering
\epsfig{file=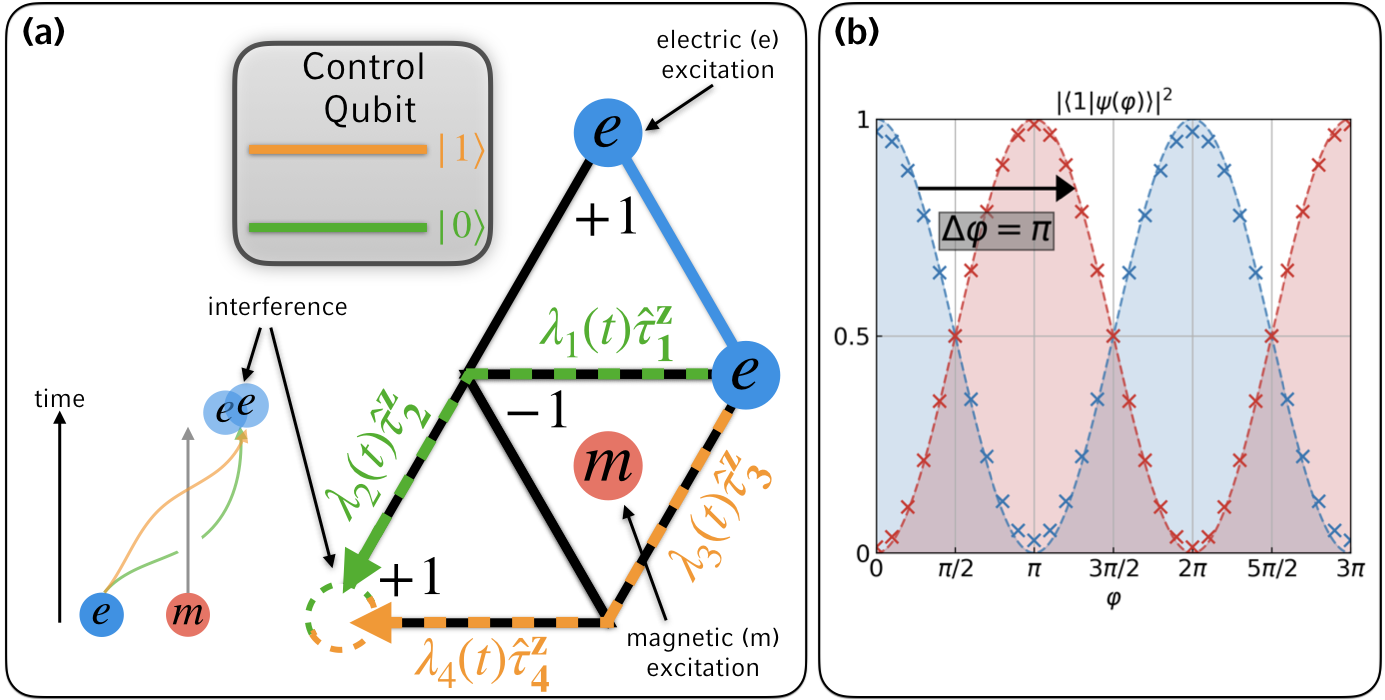, width=0.47\textwidth}
\caption{Minimal braiding scheme. (a) The braiding of excitations in real space is illustrated. First a pair of ($e$)-excitations is created (blue), which corresponds to inverting the gauge sector (or vertex terms~$\G_{V_i}$) at the position of the ($e$)-excitation. Then one of the excitations is braided along the green or orange path depending on the state of the control qubit. The implemented braiding sequence is shown on the bottom left corner. Thus, in the Ramsey protocol the control qubit is initialized in a superposition state $(\ket{0}+\ket{1})/\sqrt{2}$ by a $\pi/2$ pulse around the $x$-axis in the Bloch sphere picture. For the braiding the operators $\lambda_{i}\hat{\tau}^z_i$ are applied as consecutive $\pi$ pulses in the link variable space. After the pulse sequence the control qubit is rotated by an angle $\varphi$ in the $x-y$~plane of the Bloch sphere and then another $\pi/2$ rotates the state around the $x$-axis. A measurement of the $\ket{1}$ occupancy for different angles $\varphi$ then determines the braiding phase. Depending on the presence of a ($m$)-excitation inside the braiding loop, the Ramsey measurement will pick up a phase shift of $\pi$. (b) The plot shows the predicted curves (dashed lines) as well as ED simulations for the Ramsey protocol using an input state that was before calculated by our proposed growing scheme ($96\%$ fidelity). The red (blue) line is the curve in the presence (absence) of a ($m$)-excitation. The interferometric scheme has the advantage that it is independent of the time-evolution of the free Hamiltonian. } 
\label{fig:braiding}
\end{figure}

The toric code has electric ($e$)- and magnetic ($m$)-excitations, which correspond to vertex ($\G_{V_i} = -1$) and plaquette ($\B_{P_n} = -1$) excitations, respectively. In the string language, the excitations correspond to the open ends of strings on the physical and dual lattice.
The braiding of ($e$)-excitations in the system can be accomplished by transporting the state between different gauge sectors in a controlled manner.
We propose a dynamical braiding scheme of an ($e$)- around an ($m$)-excitation in a minimal setup with only three plaquettes. Using a Ramsey interferometry protocol, the braiding phase $e^{i\pi}$ can be experimentally extracted~\cite{Nakamura2020} while the dynamical phase is cancelled~\cite{Han2007,Atala2013, Zohar2013}.

To prepare the ($m$)-excitation, i.e.\ a state with a $\B_P=-1$ plaquette, the growing scheme proposed in Sec.~\ref{secToricCode} can be adapted. Instead of starting in the ground state, we first initialize a high-energy state, which the system then adiabatically follows into the excited toric-code manifold with a localized ($m$)-excitation (see Appendix~\ref{supp:growing_exc}). As a next step, a pair of ($e$)-excitations has to be created by flipping a link variable in the $\hat{\tau}^x$ basis, i.e.\ applying a $\pi$-pulse around the $z$-axis on the Bloch sphere. In our proposed setup, the latter can be easily applied by introducing a potential gradient between the two coupler~$(\c, \d)$ qubits.

When one of the $(e)$-excitations is moved along a path around the $(m)$-excitation, it acquires a dynamical phase from the free time evolution of the system and a geometric phase due to the anyonic nature of excitations in topological phases. In the Ramsey interferometric scheme, the dynamical phase can be cancelled such that the braiding phase can be determined. To achieve this, the path around the $(m)$-excitation is divided into two separated paths and the $(e)$-excitation runs along the paths in opposite directions, i.e.\ halfway clockwise and halfway counterclockwise as shown in Fig.~\ref{fig:braiding}. To determine the braiding phase, the phase shift between the two states, $(e)$ going clockwise $\curveplus$ and $(e)$ going counterclockwise $\curveminus$, has to be extracted. 

Hence, we propose to control which of the paths is taken by coupling the $(e)$-excitation to an external, control qubit with internal states $\ket{0}$ and $\ket{1}$ as follows:
\begin{align}
    \begin{split}
    \hat{V}_{\curveminus}(t) &= \ket{0}\bra{0}\otimes \big(\lambda_1(t)\hat{\tau}^z_1 + \lambda_2(t) \hat{\tau}^z_2\big)\\
    \hat{V}_{\curveplus}(t) &= \ket{1}\bra{1}\otimes \big(\lambda_3(t)\hat{\tau}^z_3 + \lambda_4(t) \hat{\tau}^z_4\big),
    \end{split}
    \label{eq:ramsey_operator}
\end{align}
where the functions $\lambda_i$, $i=1,..,4$ are $\pi$-pulses in the link variable space, i.e.\ $\int \lambda_i(t)dt = \pi$. The labels of link variables are defined in Fig.~\ref{fig:braiding}. 
In Appendix~(\ref{supp:braiding_scheme}), we describe a possible implementation of a minimal scheme coupled by Eq.~(\ref{eq:ramsey_operator}) to a control qubit using the \Ztwo{}~building block (Sec.~\ref{secImplementation}).

The Ramsey protocol has the following three steps: (i) A $\pi/2$-pulse initializes the control qubits in state $\left( \ket{0} + \ket{1} \right)/\sqrt{2}$. This corresponds to a rotation around the $y$-axis on the Bloch sphere. (ii) Then one $(e)$-excitation is dynamically braided around the $(m)$-excitation by the time-dependent operator $\hat{V}(t) = \hat{V}_{\curveminus}(t) + \hat{V}_{\curveplus}(t)$; to this end, a simultaneous $\pi$-pulse of $\lambda_1$ and $\lambda_3$ is followed by a simultaneous $\pi$-pulse of $\lambda_2$ and $\lambda_4$. (iii) The braiding phase can now be extracted by first rotating the control qubit by the angle $\varphi$ around the $z$-axis followed by a $\pi/2$ pulse around the $y$-axis. The occupation in state $\ket{1}$ is then dependent on the phase shift of the two paths (see Appendix~\ref{supp:ramsey}).

The plot in Fig.~\ref{fig:braiding} shows the occupation of $\ket{1}$ for different $\varphi$ in the presence and absence of the $(m)$-excitation. We performed ED simulations in the minimal setup shown in Fig.~\ref{fig:braiding} and for the same settings as in Sec.~\ref{subsecStatePrep}. With an $(m)$-excitation present, a phase shift of $\pi$ can be measured, which corresponds to the braiding phase.

The braiding phase could as well be measured by evaluating Wilson loops in a quantum projection measurement, i.e.\ in the presence of an $(m)$-excitation we could calculate Wilson loops in the $\hat{\tau}^z$ basis directly from the snapshots. The dynamical braiding scheme, however, is a step towards more advanced braiding sequences for non-Abelian anyons as needed for quantum computation~\cite{Kitaev2003}.

\section{Summary and Outlook}
\label{secSummary}
In summary, we developed an experimentally feasible building block
that implements a \Ztwo{}-gauge coupling to a dynamical matter excitation.
We observed negligible intrinsic gauge-symmetry breaking 
during experimental time scales in our numerical studies
without fine-tuning the system's parameters.
We emphasize that this building block has great potential to enable
experimental studies of \Ztwo{} LGTs coupled to dynamical matter 
in extended 1D and 2D systems by interconnecting multiple building blocks.
Moreover, the scheme is built from very basic ingredients 
-- harmonic and anharmonic oscillators -- by statically coupling them
and might therefore be applicable to a variety of experimental platforms
to enable more general quantum simulations of LGTs~\cite{Tagliacozzo2013, Wiese2013, Wiese2014, Zohar2015, Notarnicola2015, Dalmonte2016, Kasper2017, Kuno2017, Zhang2018}.

We further achieved dominating plaquette terms by introducing separate matter excitations 
on individual plaquettes of a triangular lattice.
This opens a realistic pathway for experimental investigations
of the toric-code ground state and its topological excitations.
We also provided an efficient method to grow topologically non-trivial states
and found a high preparation fidelity and a good preparation time scale
that scales proportional to $\sqrt{N}$, 
which is much faster than directly driving through the phase transition.
We studied snapshots of the charge density and concluded 
that they well serve as experimental signatures for the topological phase.
We also analyzed a minimal braiding scheme and an interferometric probe,
which makes the braiding phase accessible to experiments.
Extending the braiding scheme to non-trivial topologies, e.g.\ a pointed disk, 
may enable braiding of non-Abelian anyons
as well as the direct observation of ground state degeneracy.

\section*{Acknowledgements}
The authors would like to thank L.~Barbiero, A.~Bohrdt, N.~Goldman,  M.~Hafezi, M.~Kebric, M.~Knap, F.~Palm, F.~Pollmann for fruitful discussions. 
This research was funded by the Deutsche Forschungsgemeinschaft (DFG, German Research Foundation) via Research Unit FOR 2414 under project number 277974659, and under Germany's Excellence Strategy -- EXC-2111 -- 390814868. 
L.H.\ acknowledges support from the Max Weber Program.
C.S.\ has received funding from the European Union’s Framework Programme for Research and Innovation Horizon 2020 (2014-2020) under the Marie Sk{\ldash}odowska-Curie Grant Agreement No.\ 754388 (LMUResearchFellows) and from LMUexcellent, funded by the Federal Ministry of Education and Research (BMBF) and the Free State of Bavaria under the Excellence Strategy of the German Federal Government and the Länder.
M.A.\ acknowledges funding from the European Research Council (ERC)
under the European Union’s Horizon 2020 research 
and innovation programme (grant agreement No.\ 803047).
A.F.\ was supported by the Australian Research Council Centre of Excellence 
for Engineered Quantum Systems (EQUS, CE170100009).

\appendix
\section{Building block for \Ztwo{} LGT in 2nd-order perturbation theory}
\label{supp:2nd_order}
In the following, the second-order calculations for the effective \Ztwo{} LGT Hamiltonian are explained. The first Sec.~\ref{supp:single_block_scqubit} describes the heart of the model, which is a single building block with two matter sites and a link variable-dependent hopping. In Sec.~\ref{supp:merging_blocks} the combination of two building blocks is described. First the combination into a chain, second into a double link block. The double link block consists of four matter sites but only a single coupler. This enables the construction of the triangular lattice with hopping restricted to single plaquettes. In Appendix~\ref{supp:numerics_triangle} we underline the analytic second-order calculations with numerics for a full triangle with realistic parameters.

\subsection{Single building block}
\label{supp:single_block_scqubit}

\begin{figure}
\centering
\epsfig{file=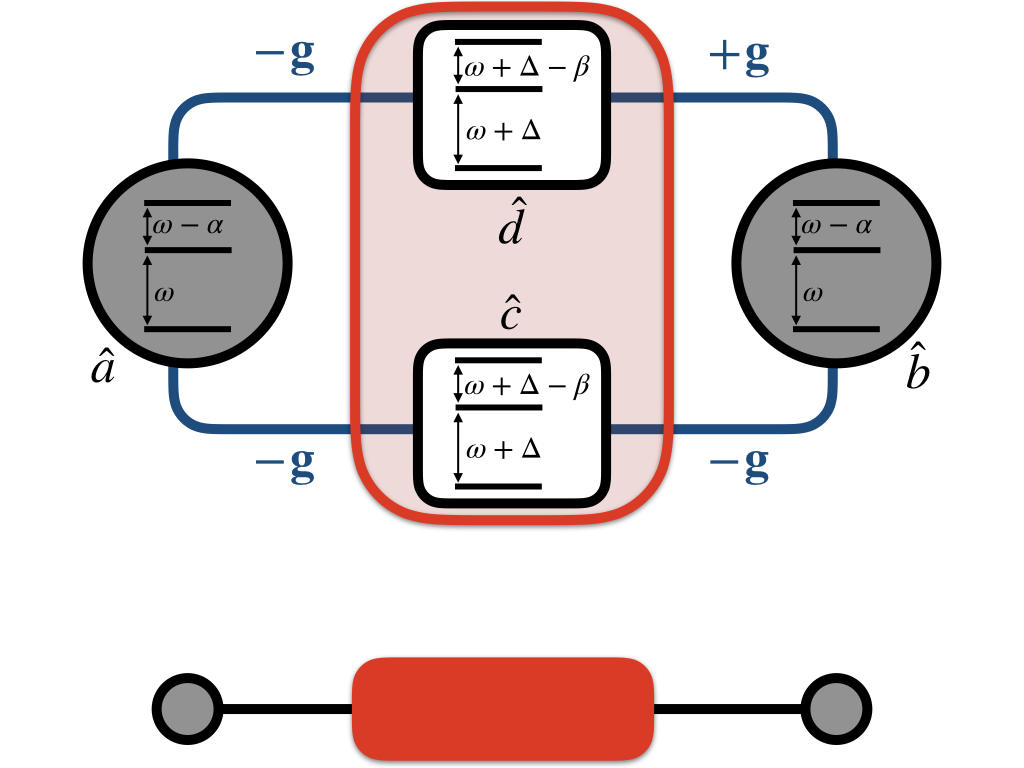, width=0.47\textwidth}
\caption{Single building block. The single building block contains two matter sites $A$ ($\a$) and $B$ ($\b$) as well a two coupler sites $C$ ($\c$) and $D$ ($\d$) that are shifted by $\Delta$ in energy compared to the matter sites. The coupler sites together define one link variable Eq.~(\ref{supp:coupler_definition}). For the beginning, we introduce anharmonicities $\alpha$ and $\beta$ whereas $\alpha \equiv 0$, ultimately.  } 
\label{fig:supplement_single_block_scqubit}
\end{figure}

We first discuss the implementation of the single \Ztwo{} building block Eq.~(\ref{eq:building_block}) consisting of four coupled (an)harmonic oscillators. As we will see later, two of the non-linearities vanish. Thus, the scheme is appealing to be implemented with superconducting qubits coupled to waveguide resonators. The positive and negative signs of the coupling elements~$g$ can be realized by coupling the qubits to the waveguide resonator at positions $\lambda$ and $\lambda/2$ of the resonator, respectively, where $\lambda$ is the wavelength of the microwave field in the resonator~\cite{Filipp2011,Kannan2020}. 
The starting point for the building block is a microscopic model of two `matter' sites $\a, \b$ and two `coupler' sites $\c, \d$ as shown in Fig.~\ref{fig:supplement_single_block_scqubit}. By applying second-order perturbation theory, we will derive an effective low-energy Hamiltonian as in Eq.~(\ref{eq:building_block}). The model is described as follows:
\begin{align}
    \H &= \H_0 + \H_\mathrm{anh} + \H_\mathrm{c} + \H_\mathrm{d} + \H_\mathrm{h} \label{supp:equation_full_sc_hamiltonian} \\
    \H_0 &= \sum_{j \in a,b,c,d} \omega_j\n_j \\
    \H_\mathrm{anh} &= -\sum_{j \in a,b,c,d}\frac{1}{2}\alpha_j \n_j(\n_j-1)\\
    \H_\mathrm{c} &= -g \big(\ad \c + \bd \c + \mathrm{H.c.}\big)\\
    \H_\mathrm{d} &= -g \big(\ad \d - \bd \d + \mathrm{H.c.}\big)\\
    \H_\mathrm{h} &= h \big( \cd \d + \mathrm{H.c.}\big),
\end{align}
where $\n_j$ is the number operator on site $j=a,b,c,d$; $\omega_j$ are the oscillator frequencies with $\omega_a=\omega_b=\omega$ and a detuning on the coupler sites $\omega_{c} = \omega_{d}=\omega+\Delta$. The anharmonicity $\alpha_j$ is chosen to be equal on the matter sites $\alpha_a = \alpha_b =\alpha$ and equal on the coupler sites $\alpha_c = \alpha_d =\beta$. The perturbation will be performed in $|g| \ll |\Delta|, |\alpha|, |\beta|$.

In the following, we will define the physical Hilbert space of the low-energy sector where $\n_a =0,1$, $\n_b =0,1$ and $\n_c + \n_d = 1$ whereas virtual couplings to doubly excited states (gapped by $\Delta$) will be taken into account. In the physical Hilbert space, we define the link variables as
\begin{align}
    \hat{\tau}^z &:= \n_c-\n_d\\
    \hat{\tau}^x &:= \cd\d + \mathrm{H.c.},
    \label{supp:coupler_definition}
\end{align}
where~$\hat{\tau}^x$ and~$\hat{\tau}^z$ represent spin-$1/2$ operators which fulfill the anticommutation relation $\{\hat{\tau}^x,\hat{\tau}^z\}=0$ on its domain. 
Moreover, we introduce the convenient notation:
\begin{align}
\begin{split}
\ket{\mathrm{A~site},&~_{\mathrm{C~site}}^{\mathrm{D~site}}~,\mathrm{B~site}}\\
\n_i=0 &\rightarrow \satisfaction{0}\\
\n_i=1 &\rightarrow \satisfaction{1}\\
\n_i=2 &\rightarrow \satisfaction{4}.
\end{split}
\label{supp:eq_notation_state}
\end{align}

\begin{table*}[p]
    \centering
    \input{Table_2nd_order_new.tex}
    \label{supp:table_2nd_order}
    \caption{Summary of the matrix elements in second-order perturbation theory for three different energy sectors. The first~$8$ rows show couplings in the singly occupied matter sector, row~$9-10$ the doubly occupied sector and row $11-12$ the states with no excitation on the matter site. The last column summarizes the second-order effective couplings in operator formalism. Row~$1-2$ are the essential terms that can be described by the Hamiltonian Eq.~(\ref{eq:building_block}). In order to cancel all undesired terms~$\propto \hat{\tau}^x$, we require rows~$5-6$ to vanish, i.e.\ by choosing~$\alpha \equiv 0$. The notation of the states follows Eqs.~(\ref{supp:eq_notation_state})}.
\end{table*}

The second-order couplings from second-order perturbation theory are summarized in Tab.~\ref{supp:table_2nd_order}. Calculations of the coupling element between a matter excitation hopping from site $A$ to $B$, show that -- by construction -- the hopping amplitude changes sign depending on the sign of $\hat{\tau}^z$. For the \Ztwo{} invariant theory, the Gauss' law $\G_{A/B} = (-1)^{\n_{a/b}} \hat{\tau}^x$ needs to be conserved. Therefore, no gauge breaking terms are allowed, which indeed cancel exactly to zero in our setting for any set of parameters~$g, \alpha, \beta$ and~$\Delta$. Thus the building block is per construction \Ztwo{}~gauge invariant in the second-order regime within the physical subsectors which have matter excitations $\n_a =0,1$ and $\n_b =0,1$. 

However, to implement the dynamics of Hamiltonian~(\ref{eq:building_block}), we require the gauge--invariant terms $\propto \n_i\hat{\tau}^x$ to vanish. These terms can be found in row $5$ and $6$ of Tab.~\ref{supp:table_2nd_order} and lead to the condition $\alpha \equiv 0$, i.e.\ no non-linearities on the matter sites $A$/$B$. The remaining terms in Tab.~\ref{supp:table_2nd_order} are dispersive energy shifts independent of the coupler sites. With dispersive energy shifts, we mean second-order processes that do not flip~$\hat{\tau}^x$ or involve charge motion. All in all, for the single building block we can conclude:
\begin{align}
    \H_{\mathrm{eff}} &= -t \big( \ad \hat{\tau}^z \b + \mathrm{H.c.} \big) + h\,\hat{\tau}^x \label{supp:eff_Ham_single_building_block}\\
    \G_{A/B} &= (-1)^{\n_{a/b}} \hat{\tau}^x  \\
    t &= 2g^2\Big(\frac{1}{\Delta - \beta} - \frac{1}{\Delta} \Big). \label{supp:eq_teff_2ndorder}
\end{align}
The effective hopping amplitude $t$ can in particular be tuned. 

For the effective Hamiltonian~(\ref{supp:eff_Ham_single_building_block}) on the single building block, we only included the sector with~$\n_a+\n_b=1$. In the full triangle, however, building blocks with~$\n_a+\n_b=0$ can occur which have a (gauge-invariant) dispersive shift compared to the sector~$\n_a+\n_b=1$. Fine-tuning of the parameters between different building blocks can cancel the dispersive shifts as explained in Sec.~\ref{supp:numerics_triangle}. The last step remaining is to verify that we can indeed glue building blocks together to construct a full triangular lattice. 

\begin{figure}
\centering
\epsfig{file=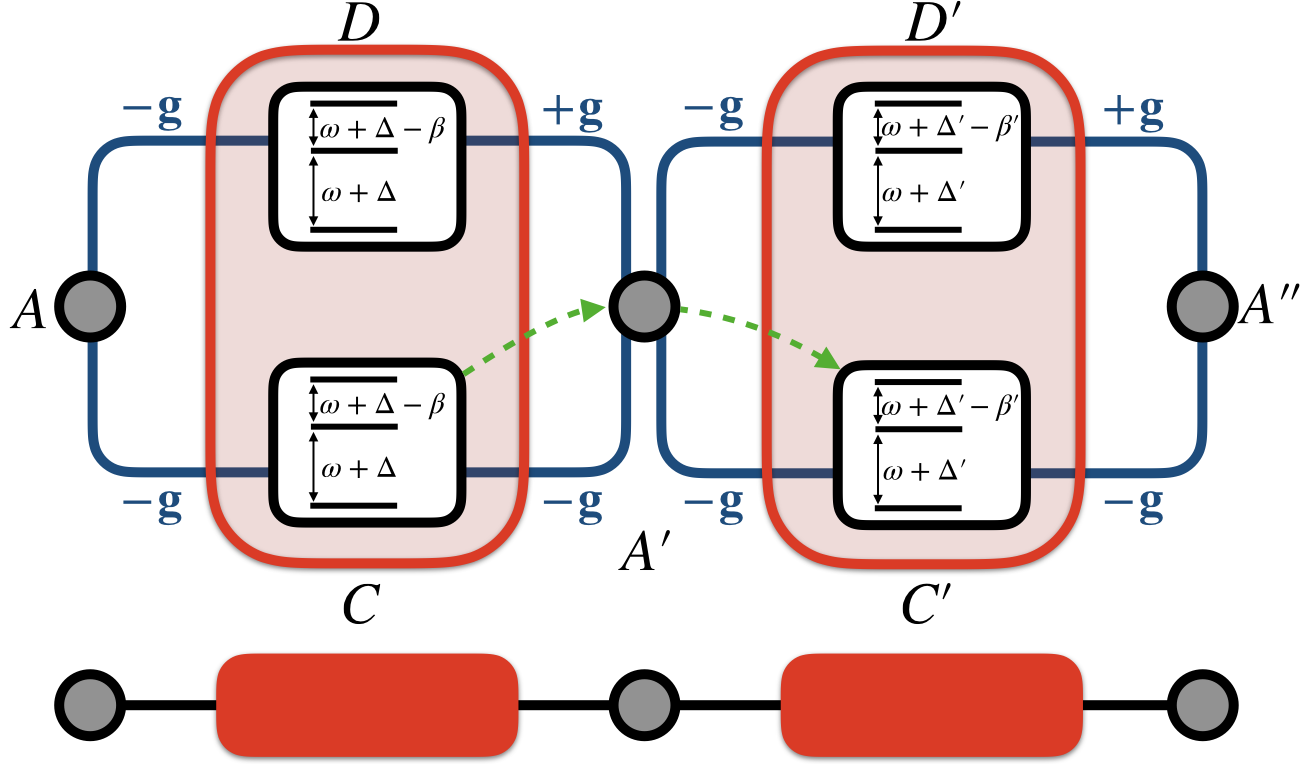, width=0.47\textwidth}
\caption{Constructing larger systems. Two single building blocks from Fig.~\ref{fig:supplement_single_block_scqubit} can be merged together such that both share the same matter site. Here, $A,A'$ and $A''$ are the matter sites and $(C,D),(C',D')$ the coupler sites. To suppress coupler-coupler hopping (depicted in green), a detuning $\delta=\Delta-\Delta'$ in the energy offsets $\Delta, \Delta'$ is introduced. The non-linearities $\beta,\beta'$ remain as free parameters to finetune the system. However, gauge symmetry is fulfilled independently.} 
\label{fig:supp_merging}
\end{figure}

\begin{figure}[b]
\centering
\epsfig{file=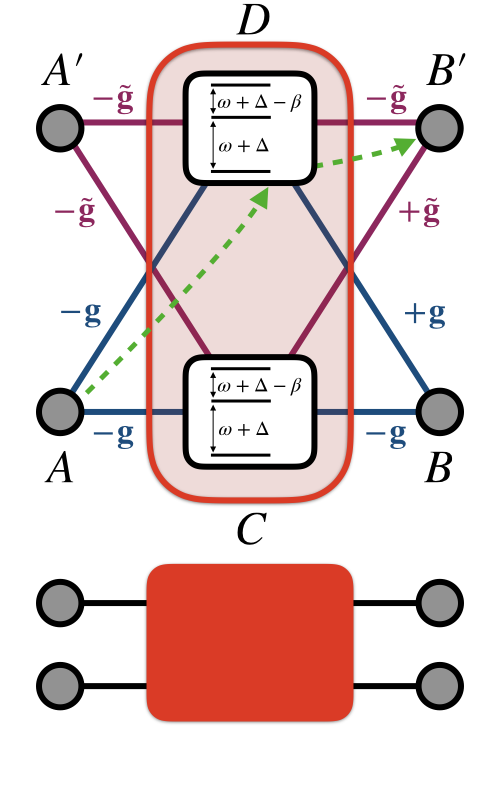, width=0.27\textwidth}
\caption{Double link. The lower part of the Figure shows the double link element as used in Fig.~\ref{fig:mic_model} and consists of two single building blocks that share the same link variable. On the top, we show the possible realization in our model. The site $A, A', B$ and $B'$ are matter sites and $C, D$ are the coupler sites. The green arrow indicates a second-order process that should be suppressed. To this end, energy offsets on the $(A',B')$ sites are introduced and the system is finetuned via the coupling parameter $\tilde{g} \neq g$. The finetuning is only necessary if homogeneous coupling $t_{A\leftrightarrow B} = t_{A'\leftrightarrow B'}$ is desired.}
\label{fig:supp_merging_H}
\end{figure}

\subsection{Merging single building blocks}
\label{supp:merging_blocks}
In Sec.~\ref{supp:single_block_scqubit}, we have discussed $A$-$B$ couplings for a single building block. To build more interesting models, as the toric code described in the main text, \Ztwo{} invariant building blocks have to be combined without creating unwanted couplings or gauge breaking terms. We will solve this problem by introducing disorder on the different coupler sites as explained now. 
Fig.~\ref{fig:supp_merging} shows two combined single building blocks with a total of three matter sites $A, A', A''$ and two double couplers $C,D$ and $C', D'$. The only relevant, new processes couple the pairs $(C,D)\leftrightarrow (C',D')$. Assuming we choose the offset $\Delta = \Delta'$, then the different couplers are on resonance and second-order perturbation theory leads to an effective coupling of $t_{(C,D)\leftrightarrow (C',D')} = \pm 2g^2/\Delta$, which is on the same order as $t_\mathrm{eff}$ (Eq.~(\ref{supp:eq_teff_2ndorder})) and violates $\n_c+\n_d = \n_{c'}+\n_{d'}=1$. To suppress this resonant coupling, we introduce a detuning $\delta := \Delta - \Delta'$ between the two coupler pairs. In the limit where $ |g^2/\Delta| \ll |\delta|$, the effective Rabi coupling between the pairs becomes $t^\delta_{(C,D)\leftrightarrow (C',D')} = \pm g^4/(\Delta^2\delta)$ and thus negligible.

By introducing disorder potentials~$\Delta_i$ on the coupler sites~$i$, we can engineer a variety of \Ztwo{} LGTs from the single building blocks. In order to ensure equal couplings~$t$ in the effective Hamiltonian, we propose to also choose disorder on the anharmonicities~$\beta_i$. After fixing the disorder potentials~$\Delta_i$ and the desired couplings~$t$, the anharmonicities~$\beta_i$ can be chosen as per Eq.~(\ref{supp:eq_teff_2ndorder}).

\begin{figure*}
\centering
\epsfig{file=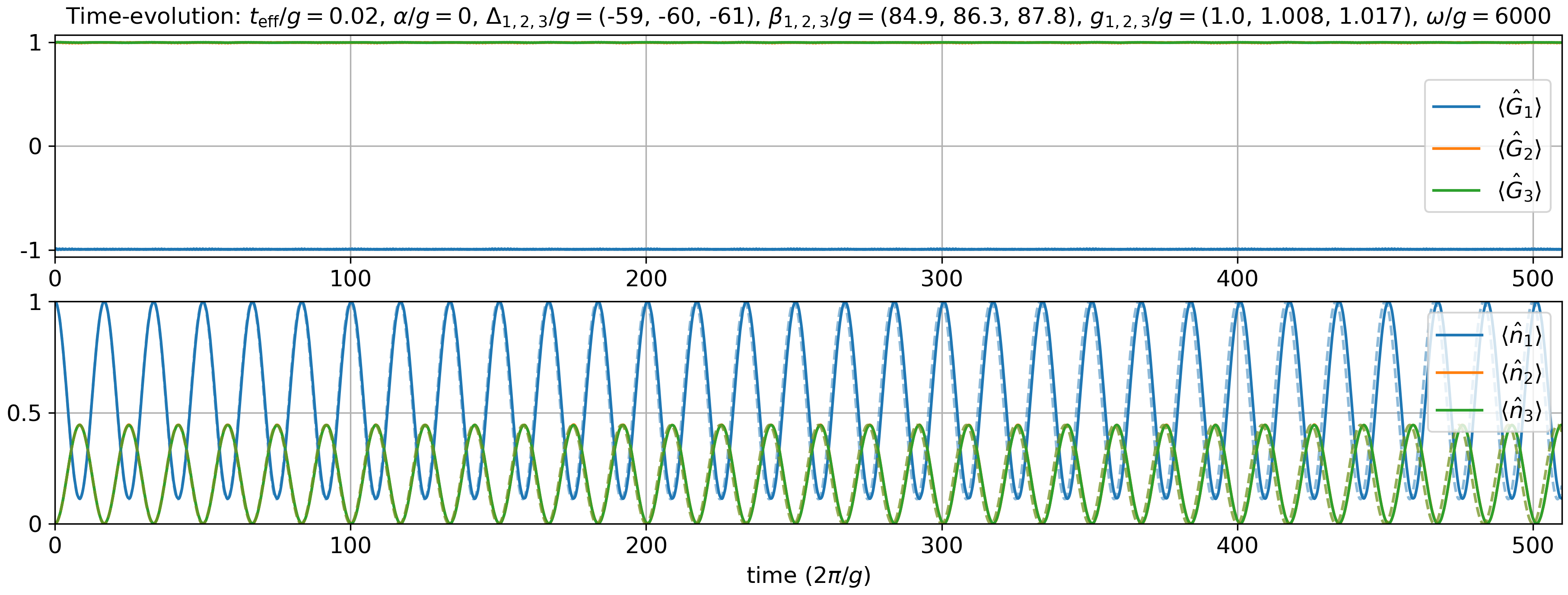, width=\textwidth}
\caption{Numerical simulation for full triangle. The plots show the time-evolution using exact diagonalisation for a full triangle with disorder in~$\Delta$'s an~$\beta$'s (see~Figs.~\ref{fig:mic_model}a and~\ref{fig:supp_merging}) and local Hilbert space dimension of ${d_{\text{max}}=3}$. The initial state has a matter excitation sitting on site~$1$ and link variables being in the $\hat{\tau}^x=+1$ eigenstate. In the upper plot the expectation values $\G_i$ on site~${1,2,3}$ are shown. They are conserved during the time-evolution, which shows the conservation of Gauss' law. The lower plot shows the site occupation of the matter excitations versus time. Here, the solid line corresponds to the microscopic second-order model and the pale dashed line is a calculation for the toy model with the same effective hopping $t_\mathrm{eff}/g = 0.02$. In both plots the orange curve is covered by the green curve. }
\label{fig:supp_numerics_triangle}
\end{figure*}

Finally, we argue that the double link, i.e.\ two single building blocks connected via the coupler sites as used in Fig.~\ref{fig:mic_model}, is well-defined. The coupling scheme we propose for the double link is shown in Fig.~\ref{fig:supp_merging_H}. The relevant processes to be suppressed are couplings between $(A,B)$ and $(A',B')$ sites. To this end, the pairs $(A,B), (A',B')$ have to be shifted out of resonance which can be accomplished by adding an energy offset $\tilde{\delta}$ onto the sites $(A',B')$. Following the argument from the previous paragraph, the unwanted coupling now becomes $\pm 2g\tilde{g}/(\Delta^2 \tilde{\delta})$, where $\tilde{g}$ is defined as in Fig.~\ref{fig:supp_merging_H}. Since the offset $\Delta$ and anharmonicity $\beta$ on the $(C,D)$ sites are fixed, the effective interaction strength between $A'\leftrightarrow B'$ will change after introducing the energy offset $\tilde{\delta}$. Now, the last free parameter $\tilde{g}$ can be used to engineer a homogeneous coupling $t_{A\leftrightarrow B} = t_{A'\leftrightarrow B'}$ across the whole system according to Eq.~(\ref{supp:eq_teff_2ndorder}).

\subsection{Numerical simulation for full triangle}
\label{supp:numerics_triangle}
To underline our predictions from Sections~\ref{supp:single_block_scqubit} and~\ref{supp:merging_blocks}, we performed an exact diagonalisation analysis of the microscopic Hamiltonian~(\ref{supp:equation_full_sc_hamiltonian}) and compared it to the ideal $\mathbb{Z}_2$ model Eq.~(\ref{eq:building_block}). For this, we calculate two quantities: the Gauss' laws $\G_i$ and the dynamic of the matter excitations as shown in Fig.~\ref{fig:supp_numerics_triangle}.
A full triangular configuration was initialized with a single matter excitation on site $1$ and all link variables pointing along $\tau^x=+1$ before the system was time-evolved. The parameters were chosen such that an effective hopping $t_\mathrm{eff}/g = 0.02$ according to Eq.~(\ref{supp:eq_teff_2ndorder}) for fixed $\beta_i$ is achieved, where~$i=1,2,3$ labels the three sites. The $\beta_i$ and corresponding~$\Delta_i$ have a slight disorder $|\delta/g| = 1 \ll |2g^2/\Delta|$ as proposed in Sec.~\ref{supp:merging_blocks}. Due to the disorder the system has different dispersive energy shifts depending on the position of the matter excitation (cf. Tab.~\ref{supp:table_2nd_order}). To enable a direct comparison with the ideal model, we finetuned hoppings $g_i$ on the links in the following way, which solves for equal shifts and equal effective hoppings on all links:
\begin{itemize}
    \item Set $\dfrac{g_1}{g}=1$
    \item Choose $\dfrac{t_\mathrm{eff}}{g}$ and $\dfrac{\beta_{1,2,3}}{g}$
    \item Calculate $\dfrac{\Delta_1}{g} = \dfrac{\beta_1}{g}\left( \dfrac{1}{2} - \sqrt{1 + \dfrac{8g_1^2}{t_\mathrm{eff}\beta_1}} \right)$
    \item The remaining parameters follow by solving the constraints: $\dfrac{\Delta_{2,3}}{g} = \dfrac{\beta_{2,3}\Delta_1}{\beta_1 g}$ and $\dfrac{g_{2,3}^2}{g^2} = \dfrac{\beta_{2,3}}{\beta_1}$
\end{itemize}
The Gauss' laws are very well conserved for any duration the system was evolved, which shows that we indeed implement a \Ztwo{} LGT. Also the dynamics of the matter excitation only show tiny deviations after long times, which is presumably due to a slightly different $t_\mathrm{eff}$ from higher order contributions. Note, however, the Gauss' law is fulfilled independent of this finetuning procedure.

\section{Derivation of effective Hamiltonian}
\label{supp:eff_hamiltonian}
Here we derive the effective Hamiltonian with multiple plaquettes and give the explicit form of the basis transformation on the full $2$D lattice. We start from the general free ($h=0$) theory represented by the Hamiltonian with multiple plaquettes~(\ref{eq:mic_hamiltonian})
\begin{align}
    \H = -t\sum_{n}\sum_{\lrangle{i}{j}\in P_n} \Big(\adn_i \tauz{i}{j} \an_j + \mathrm{H.c.}\Big),
\end{align}
where the hopping with amplitude $t\,\tauz{i}{j}$ is restricted within a single plaquette $P_n$. The number of matter excitations $\a^{(n)}$ per plaquette $P_n$ is exactly one boson. The goal is to distribute the acquired, local phases equally among all the hopping terms within a plaquette by exploiting a gauge transformation acting on the $\an_i$'s. The transformation $\U$ is a straight forward extension of the single plaquette transformation~(\ref{eq:gauge_trafo_explicit}): 
\begin{align}
    \Ud \an_j \U &= \an_j e^{i\hat{\vartheta}_j[\hat{\tau}^z]}\\
    \label{eq:supp_gauge_trafo_U}
    \U &= e^{i\sum_n \sum_j \hat{\vartheta}^{(n)}_j[\hat{\tau}^z]\adn_j \a^{(n)}_j}\\
   \hat{\vartheta}^{(n)}_j[\tau^z] &=  \frac{\pi}{2}\left(\hat{\tau}^z_{\langle j,j+1 \rangle_n} - \hat{\tau}^z_{\langle j-1,j \rangle_n}\right)
\end{align}
    \begin{align}
    \Ud \H \U = & -t \sum_n \sum_{\lrangle{i}{j}\in P_n}\Big\{\\ &\adn_i \hat{\tau}^z_{\langle i-1,i \rangle_n}\hat{\tau}^z_{\langle i,i+1 \rangle_n}\hat{\tau}^z_{\langle i+1,i+2 \rangle_n} \an_j \nonumber\\ &\quad+\mathrm{H.c.}\nonumber\Big\},
    \end{align}
where we can define the plaquette operator $\B_{P_n} = \hat{\tau}^z_{\langle i-1,i \rangle_n}\hat{\tau}^z_{\langle i,i+1 \rangle_n}\hat{\tau}^z_{\langle i+1,i+2 \rangle_n}$ for $\langle \cdot , \cdot \rangle_n \in P_n$. In momentum representation $\a_j^{(n)} = 3^{-1/2} \sum_{k_m=\frac{2\pi}{3}m}e^{ikR_i}\a_{k_m}^{(n)}$ (lattice spacing is set to $\delta=1$), the Hamiltonian can be written as in Eq.~(\ref{eq:hamiltonian_U}). 

The next claim is that the local symmetry generators $\G_i = (-1)^{\hat{N}_i}\prod_{j:\ij}\taux{i}{j}$ of the theory, resemble the vertex operators $\G_{V_i} = \prod_{j:\ij}\taux{i}{j}$ of the toric code in the transformed basis $\U$. For this, we consider the transformation of a single $\taux{i}{j}$ term. By explicitly expanding the exponential, the transformation rule can be derived:
\begin{align}
    \begin{split}
    &\Ud \taux{i}{j} \U = e^{-i\frac{\pi}{2}\tauz{i}{j}\Delta\n} \taux{i}{j} e^{i\frac{\pi}{2}\tauz{i}{j}\Delta\n}\\
    &= \left[ \cos{\left( \frac{\pi}{2}\Delta\n \right)} - i\tauz{i}{j}\sin{\left( \frac{\pi}{2}\Delta\n \right)} \right]\\
    &\times\taux{i}{j}\,\left[ \cos{\left( \frac{\pi}{2}\Delta\n \right)} + i\tauz{i}{j}\sin{\left( \frac{\pi}{2}\Delta\n \right)} \right],\label{supp:eq_taux_trafo}
    \end{split}
\end{align}
where $\Delta\n$ is the difference of matter excitation between the two ends of the link $\ij$, where only matter excitations count that are directly attached to the link. On the double link (Fig.~\ref{fig:mic_model}b) this corresponds to the imbalance of matter excitations on the different sides of the link variable. Since we have restricted the number of matter excitations to strictly one per plaquette, the eigenvalues of $\Delta\n$ can only take the values $\Delta n = 0,\pm1,\pm2$. Thus, the transformation of the \Ztwo{}~electric field is
\begin{align}
    \Ud \taux{i}{j} \U = (-1)^{\Delta\n}\taux{i}{j}.
    \label{eq:supp_taux_trafo}
\end{align}

Now, we can calculate the transformation of the symmetry generators $\G_i$:
\begin{align}
\begin{split}
    \Gt_i &= \Ud \G_i \U = (-1)^{\hat{N}_i} \prod_{j: \lrangle{i}{j}} \Ud \taux{i}{j} \U \\
    &= (-1)^{\sum_{i: P_i}\hat{n}_{P_i}}\prod_{j: \lrangle{i}{j}}\taux{i}{j}
\end{split}
\end{align}
Here, $\sum_{i: P_i}\hat{n}_{P_i}$ is the sum of matter excitations over all plaquettes connected to vertex $i$. Since we work in a sector, where the number of excitations per plaquette is restricted to strictly one, the Gauss' laws simplify to
\begin{align}
     \Gt_i = \Ud \G_i \U &= (-1)^{N^P_i}\prod_{j: \lrangle{i}{j}}\taux{i}{j}\\
     &=(-1)^{N^P_i}\G_{V_i},
\end{align}
where $N^P_i$ are the number of plaquettes connected to vertex~$i$. Therefore, in the transformed basis the Gauss' laws are given by the vertex operators $\G_{V_i}$ up to a fixed pre-factor.

\begin{figure}
\centering
\epsfig{file=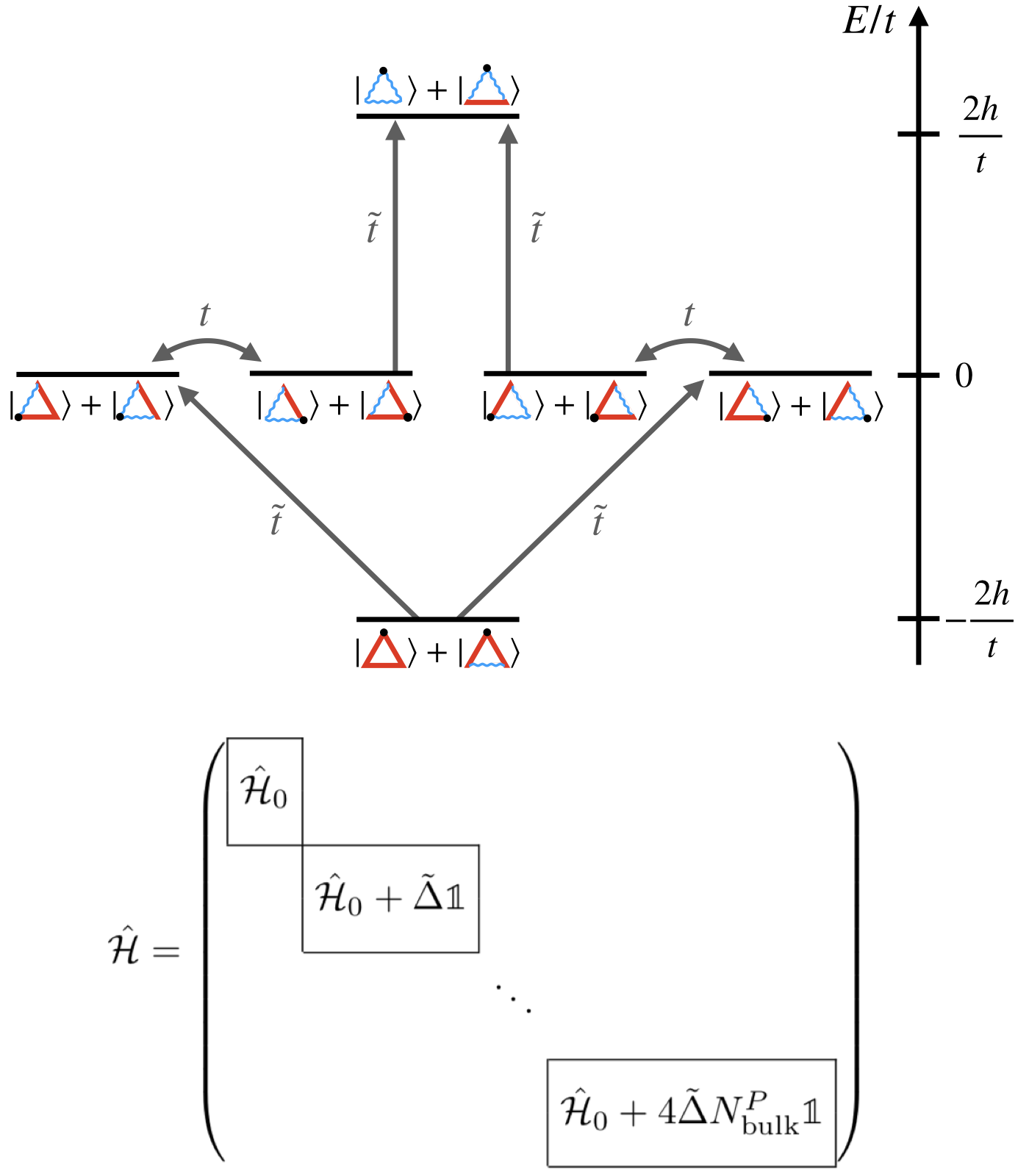, width=0.47\textwidth}
\caption{Growing scheme -- Induction step. In the basis described in Fig.~\ref{fig:growing_scheme}b, the  Hamiltonian~(\ref{eq:growing_hamiltonian}) can be written in block matrix form with blocks of size~$(6\times6)$ and energy sectors are separated by a gap~$\tilde{\Delta}$ which is the energy gap between excitation sectors in the bulk. The couplings of an individual block are shown on the top. The local Gauss' laws restrict the Hilbert space dimension to~$3\cdot2^3/2^2=6$. } 
\label{fig:supplement_growing_scheme}
\end{figure}
\section{State preparation and manipulation}
\label{supp:state_prep_mani}
In Appendix~\ref{supp:growing_scheme} details about the growing schemes together with analytical and numerical arguments are explained. Appendix~\ref{supp:hidden_order} discusses the transformation rules of the snapshots under $\U$ in order to reveal the closed loop configurations of strings in the toric-code ground state. In Appendix~\ref{supp:growing_exc} and~\ref{supp:ramsey} we calculate the preparation of local, magnetic flux excitations in the system as well as the theoretical predictions of the Ramsey scheme.
\subsection{Growing scheme}
\label{supp:growing_scheme}

In this section, we show that for every individual growing step a path with constant gap~$\Delta=t$ exists. We consider the setup described in Fig.~\ref{fig:growing_scheme}a, where the bulk in the topological phase and a single plaquette on the boundary is in its trivial phase. Since the \Ztwo{} electric field term~$h$ in the bulk is turned off, the bulk Hamiltonian can conveniently be described in the transformed basis (see Appendix~\ref{supp:eff_hamiltonian})
\begin{align}
    \U_\mathrm{bulk} &= e^{i\sum_{n \in \mathrm{bulk}} \sum_j \hat{\vartheta}^{(n)}_j[\hat{\tau}^z]\adn_j \an_j},
\end{align}
which is a transformation that only acts on the matter sites in the bulk. The Hamiltonian in the new basis~$\U_\mathrm{bulk}$ is then given by
\begin{widetext}
    \begin{align}
    \begin{split}
	\Ud_{\mathrm{bulk}}\H \U_{\mathrm{bulk}} &= -t\sum_{n\in\mathrm{bulk}}\sum_{k_m}\cos{(k_m + \hat{\Phi}^{(n)})}\adn_{k_m}\an_{k_m}  \\ &+ (-t\ad_1\tauz{1}{2}\a_2-\tilde{t}\ad_2\tauz{2}{3}\a_3 -\tilde{t}\ad_3\tauz{3}{1}\a_1 + \mathrm{H.c.} )
	+ h\taux{2}{3} + h\taux{3}{1}, 
    \label{eq:growing_hamiltonian}
    \end{split}
    \end{align}
\end{widetext}
where the parameters $\tilde{t}$ and $h$ are the tunable parameters in the adiabatic scheme, while $t$ is fixed. The phase shift $\hat{\Phi}^{(n)}$ is defined as in Eq.~(\ref{eq:phase_plaquette}) and depends on the plaquette operators $\B_{P_n}$. The labels of the link variables and matter sites are defined in Fig.~\ref{fig:growing_scheme}a. Since the first and second line of Eq.~(\ref{eq:growing_hamiltonian}) commute, the operators which are used for the growing step decouple from the bulk states. Therefore, the bulk basis states can be reduced to only the ground state per assumption. The physically relevant Hilbert space is thus spanned by the bulk ground state and the states of the single plaquette, which have dimension~$3\cdot2^3/2^2=6$ due to the local Gauss' law constraints. Hamiltonian~(\ref{eq:growing_hamiltonian}) can therefore be written by $(6\times6)$~block matrices and in Fig.~\ref{fig:supplement_growing_scheme} one such six-level scheme with all couplings is depicted. The individual blocks are shifted in energy by $\Delta=t$ as shown in Fig.~\ref{fig:supplement_growing_scheme}.

\begin{figure}[b]
\centering
\epsfig{file=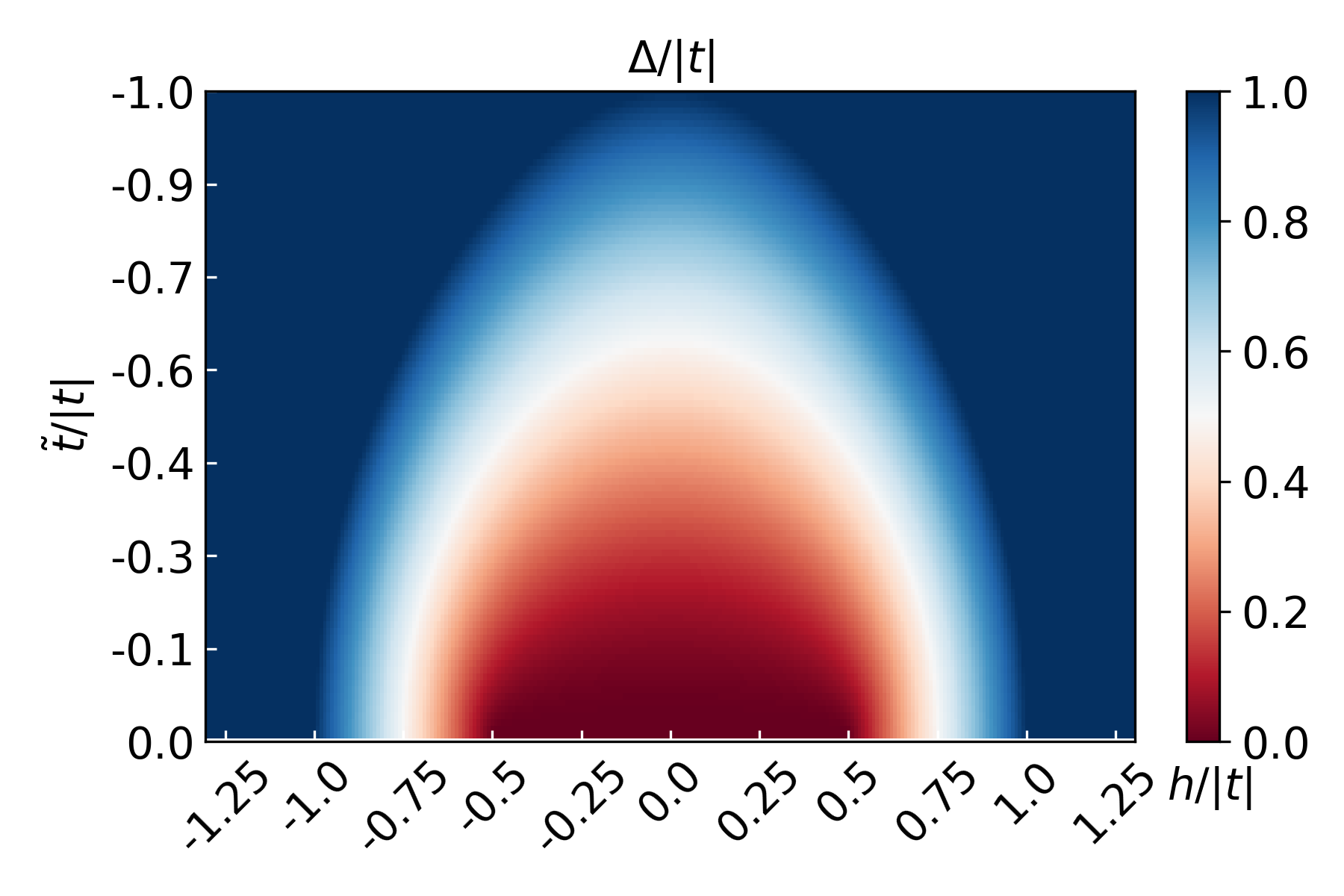, width=0.47\textwidth}
\caption{Growing scheme -- Analytical gap calculation. The plot shows the gap $\Delta$ between the ground and excitated state manifolds depending on the tunable parameters $\tilde{t}$~and $h$~for a single growing step. The results are obtained by diagonalizing $(6\times 6)$~matrices. The gap landscape has the same shape as in our ED analysis shown in Fig.~\ref{fig:growing_scheme}f. }
\label{fig:supplement_growing_gap}
\end{figure}

The $(6\times6)$~matrices can be easily diagonalized using numerical methods or computer algebra programs. In Fig.~\ref{fig:supplement_growing_gap}, we plot the gap $\Delta$ between the ground state and first excited state of the many-body spectrum versus the tunable parameters $\tilde{t}$ and $h$. The plot resembles the gap landscape shown in Fig.~\ref{fig:growing_scheme}f, which was derived in an ED calculation on a lattice with three plaquettes. The parameter path suggested in the main text therefore has a constant energy gap $\Delta=t$.

The above arguments hold specifically when the system has a bulk and a single plaquette on the boundary. However, other configurations can also appear as in Fig.~\ref{fig:growing_scheme}d, where the matter excitation is initially delocalized on the sites~$\a_1$ and~$\a_2$. The system can still be described by Hamiltonian~(\ref{eq:growing_hamiltonian}) but only the gauge sector changes. The essential argument and calculation is valid for all configurations that can appear on the triangular lattice.

Therefore, the very generic argument shows that for each growing step, there always exists a parameter path with a constant gap $\Delta=t$ connecting between the trivial and topological phase.

\subsection{Hidden topological order}
\label{supp:hidden_order}
The toric-code ground state, which is a superposition of all closed loop configurations, is the ground state of Hamiltonian~(\ref{eq:hamiltonian_U}) in the transformed basis. Projective measurements (snapshots), however, are taken in the laboratory frame. The transformation~$\U$, that depends on the local densities~$\adn_i \an_i$, has to be evaluated to go from the old into the new basis. Since the link variables~$\taux{i}{j}$ do not commute with~$\U$, they have non-trivial transformation laws~(\ref{supp:eq_taux_trafo}) which can be summarised as follows:
\begin{itemize}
    \item Choose a link variable $\taux{i}{j}$.
    \item Calculate the imbalance $\Delta n$ between matter excitations on site~$i$ and~$j$ of the link $\ij$. Only take into account matter sites that are directly attached to the link variable $\taux{i}{j}$.
    \item The sign of the link variable flips iff~$\Delta n$ is odd. This corresponds to the creation/annihilation of a string in the $\hat{\tau}^x$~basis.
\end{itemize}

\subsection{Growing a magnetic excitation}
\label{supp:growing_exc}
\begin{figure}[t]
\centering
\epsfig{file=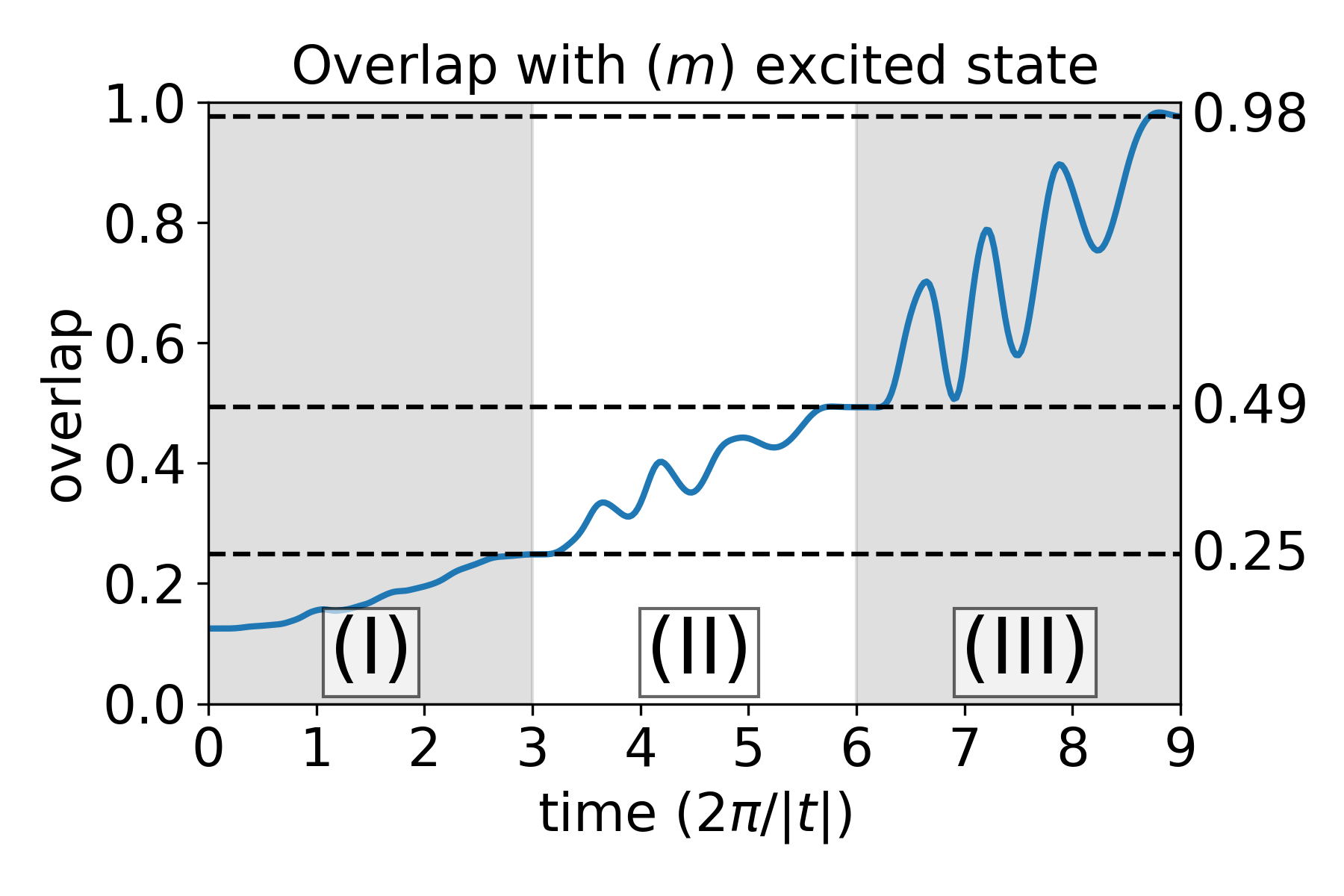, width=0.47\textwidth}
\caption{Growing a magnetic excitation. We plot the overlap $\tr_\mathrm{links}(\tr_\mathrm{matter}( \ket{\tilde{\psi}(t)}\bra{\tilde{\psi}(t)})\ket{\Psi_\mathrm{m}}\bra{\Psi_\mathrm{m}})$ between the time-evolved state $\ket{\tilde{\psi}(t)}$ and the toric-code state $\ket{\Psi_\mathrm{m}}$ with a magnetic ($m$)-excitation. The excitation is placed on the center plaquette in the three plaquette configuration (see Fig.~\ref{fig:braiding}). Section (I)-(III) describes the growing of plaquette $P_1$ to $P_3$. For the chosen timescale, the growing scheme shows a large overlap~$>98\%$.} 
\label{fig:supp_growing_mag_exc}
\end{figure}

Excitations in the toric-code are known to have anyonic statistics. To study this physics, we need to prepare a state with a well-defined excitation, i.e.\ a localized~$\B_P=-1$ term (vison). To realise this in our setup, we adapt the ground-state preparation scheme. Instead of following the ground state of the Hamiltonian, we initialize the system in an excited eigenstate. Therefore, during the growing step the system follows its eigenstate into a toric-code eigenstate with a localized $(m)$-excitation. We choose the highest-excited state in the single plaquette spectrum (see Fig.~\ref{fig:mic_model}e). The growing scheme then works completely analogously, i.e.\ our suggested parameter path for the ground-state preparation also maintains a large gap in the excited-state preparation. In principle, the excited state manifolds of the system are degenerate. However, throughout the growing scheme the system does not couple to the other excited states.

A small system ED study should underline the efficiency of the growing scheme for preparing excited states. For the minimal example of three plaquettes, we can prepare a single $(m)$-excitation on the center plaquette. Typically excitations in the toric code appear in pairs but due to the boundary condition, we can place the second excitation `outside of the lattice'. If the second excitation would additionally be located within the braiding loop, the wavefunction would pick up twice the braiding phase, i.e.\ $2\pi$, and thus could not be distinguished from the zero phase shift for no~$(m)$-excitation.

To this end, we initialize and time-evolve the system as described in Sec.~\ref{subsecStatePrep} but during the growing procedure, we flip the sign of the external field~$h$ on the link variable which is located on the boundary of the center plaquette. To quantify the fidelity of the growing scheme numerically, we -- analogously to the main text -- calculate the overlap with a toric-code state that has a localized ($m$)-excitation $\tr_\mathrm{links}(\tr_\mathrm{matter}( \ket{\tilde{\psi}(t)}\bra{\tilde{\psi}(t)})\ket{\Psi_\mathrm{m}}\bra{\Psi_\mathrm{m}})$ as shown and defined in Fig.~\ref{fig:supp_growing_mag_exc}.

\subsection{Implementation of the braiding scheme}
\label{supp:braiding_scheme}
\begin{figure}
\centering
\epsfig{file=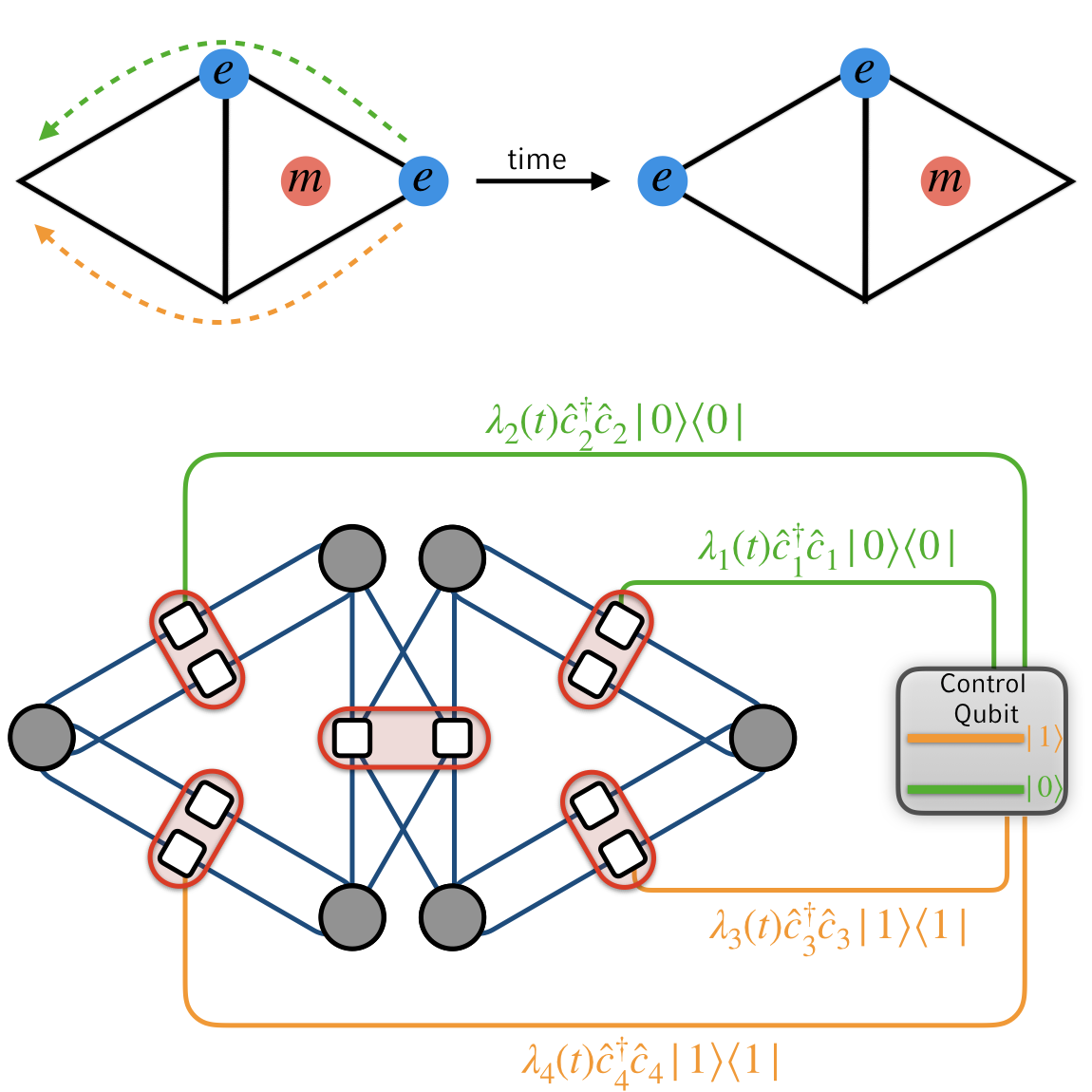, width=0.47\textwidth}
\caption{Implementation of the braiding scheme. The upper part illustrates the braiding protocol for two triangles with its initial and final state as well as the position of excitations as defined in Fig.~\ref{fig:braiding}. Whether the excitation moves along the green or orange path, is determined via the internal state of an external control qubit. The lower part shows the couplings between its microscopic constituents, e.g.\ superconducting qubits, where the individual building blocks have the same structure as in Fig.~\ref{fig:supplement_single_block_scqubit}-\ref{fig:supp_merging_H}. The external control qubit is coupled dispersivly to four single~$c$ sites (orange and green lines), which -- under the application of an additional energy tilt between the $c$ and $d$ site -- effectively yields a $\hat{\tau}^z$ term on each link required to transport the~$(e)$-excitation. Whether the interaction is sensitive to the occupation of~$|0\rangle$ (green) or~$|1\rangle$ (orange) of the control qubit can be controlled by an external potential Eq.~(\ref{eq:supp_compPot}). } 
\label{fig:supp_braiding_implement}
\end{figure}

In Sec.~\ref{secTopExcBraiding}, we propose a scheme to extract the geometric braiding phase 
from dynamically braiding an $(e)$-excitation around an $(m)$-excitation.
To this end, we use an external control qubit, that controls the two braiding paths
and is ultimately used to extract the relative phase shift between them.
In this section, we explain a realistic, minimal setup 
to implement the braiding protocol with superconducting qubits.

In the main text, we show a setup with three plaquettes for educational purposes
as it is the minimal example where one $(e)$-excitation is in the braiding loop enclosing the $(m)$-excitation.
In general, both $(e)$- and $(m)$-excitations can only come in pairs.
However, in a finite size system open boundary conditions allow 
to store one $(m)$-excitation ``outside'' of the system.
Thus, in Sec.~\ref{secTopExcBraiding}, 
we placed one $(m)$-excitation in the center and the other one ``outside''; 
in addition, we added a third plaquette not enclosed by the braiding loop, 
on which the non-participating $(e)$-excitation rests during the entire braiding scheme.

Here, we only use two plaquettes in the following discussion,
which makes it even more appealing for experimental realization due to its smaller system size.
As a consequences, the non-participating $(e)$-excitation is located within one of the braiding paths 
as shown in Fig.~\ref{fig:supp_braiding_implement}, which does not alter the physics of the participating $(e)$-excitation.
In summary, the smallest system to observe braiding in an experiment consists of only two plaquettes 
coupled to one external control qubit 
and thus six harmonic oscillators (matter sites) 
and eleven anharmonic oscillators (link variables and control qubit) are required.
An ED study of the effective model with two plaquettes is compared to the three plaquette scheme, 
which both yield the same result after applying our proposed Ramsey protocol.

The most important ingredient for the proposed scheme is to derive an implementation for the following interaction (Eq.~\ref{eq:ramsey_operator}):
\begin{align}
    \hat{V}(t) = \lambda(t) \big( \sum_{j=1,2} \hat{\tau}^z_j\otimes |0\rangle\langle0| + \sum_{j=3,4} \hat{\tau}^z_j\otimes |1\rangle\langle1|  \big), \label{eq:supp_Vt}
\end{align}
where $|0\rangle$, $|1\rangle$ are the states of the control qubit and the labels of links $j=1,..,4$ are defined as in Fig.~\ref{fig:supp_braiding_implement}.
In the \Ztwo{}~building block, the link variable is defined as~$\hat{\tau}^z =\cd\c -\dd\d$.
We can insert this identity into Eq.~(\ref{eq:supp_Vt}) by further using $\n_c+\n_d=1$, which yields
\begin{align}
\begin{split}
    \hat{V}(t) = \lambda(t) \Big(&\sum_{j=1,2} 2\cd_j\c_j \otimes|0\rangle\langle0| \\+ &\sum_{j=3,4} 2\cd_j\c_j \otimes|1\rangle\langle1| \Big) 
    + \mathrm{const.}
\end{split}
\end{align}
We can see that the operator~$\hat{V}(t)$, that transports the $(e)$-excitation, requires a density-density coupling between the control qubit and the individual $c$ sites.
We suggest to realize this interaction by dispersivly coupling the qubits as shown in Fig.~\ref{fig:supp_braiding_implement}.

For the dispersive coupling, we assume that the control qubit has a larger frequency than the $c$ sites.
Since the control qubit and the $c$ site are off-resonant, the excitations can only virtually tunnel between the two qubits leading to a dispersive energy shift on the corresponding sites.
In second-order perturbation theory, where we denote the effective coupling strength conveniently by~$\lambda(t)$ as above, we can derive the following interaction:
\begin{align}
\begin{split}
    &\hat{V}^{(2)}(t) = \lambda(t) \Big( \sum_{j=1,..,4} \cd_j\c_j\otimes|0\rangle\langle0| -\dd_j\d_j\otimes|1\rangle\langle1|\big) \\
    &=\lambda(t) \big(\sum_{j=1,..,4} \cd_j\c_j\otimes|0\rangle\langle0| +\cd_j\c_j\otimes|1\rangle\langle1| \Big) - \mathbb{1}\otimes|1\rangle\langle1| \label{eq:supp_Vt_2nd}
\end{split}
\end{align}
The terms describe the dispersive shifts on the qubit pair ($c$, $d$) with reversed sign of the coupling for the two configurations of the control qubit, which arises from the different relative energy shift between the initial/final and virtual state for the different configurations of $|0\rangle$ and $|1\rangle$.
However, Eq.~(\ref{eq:supp_Vt_2nd}) differs from the desired interaction Eq.~(\ref{eq:supp_Vt}) since the effective scheme still couples all $c$ sites to both states $|0\rangle$ and $|1\rangle$.
By applying a compensation potential~$\delta\hat{V}(t)$, which are specific detunings on the $c$ sites, we can accomplish to resemble interaction Eq.~(\ref{eq:supp_Vt}) up to a constant energy shift.
The implemented effective interaction during the braiding scheme is thus given by
\begin{align}
    \hat{V}(t) &= \hat{V}^{(2)}(t) + \delta\hat{V}(t) + \mathrm{const.} \\
    \delta\hat{V}(t) &= \lambda(t) \big(\sum_{j=1,2} \cd_j\c_j\otimes \mathbb{1} - \sum_{j=3,4} \cd_j\c_j\otimes \mathbb{1} \big).\label{eq:supp_compPot}
\end{align}

The interaction potential~$\hat{V}(t)$ has to be applied for a certain pulse time~$T$, for which the optimal time~$T$ is derived in Appendix~\ref{supp:ramsey}.
Furthermore, we require the interaction to switch on and off in a sufficiently short time; especially during state preparation the interaction to the control qubits should be absent since~$\hat{V}(t)$ intentionally couples different gauge sectors.
The fast switching can be achieved by tuning the frequency of the $c$ site such that all dispersive shifts are identically compensated.
To this end, we introduce another compensation potential $\delta\hat{V}^{\mathrm{off}}(t)$ with
\begin{align}
    &\delta\hat{V}^{\mathrm{off}}(t) = -\lambda(t)\sum_{j=1,..,4} \cd_j\c_j \otimes \mathbb{1} \\
    &\Rightarrow \hat{V}(t) + \delta\hat{V}^{\mathrm{off}}(t) = \mathrm{const.}
\end{align}

To conclude, we have presented a possible coupling scheme on the level of the microscopic constituents of the \Ztwo{}~building block.
In this minimalistic setup only two plaquettes and one additional control qubit
together with an adapted switching scheme is needed to implement the Ramsey protocol.
The initialization and detection of the braiding phase is the same as described in the main text 
and requires only single qubit rotations of the control qubit.

\subsection{Ramsey interferometry}
\label{supp:ramsey}
\begin{figure}
\centering
\epsfig{file=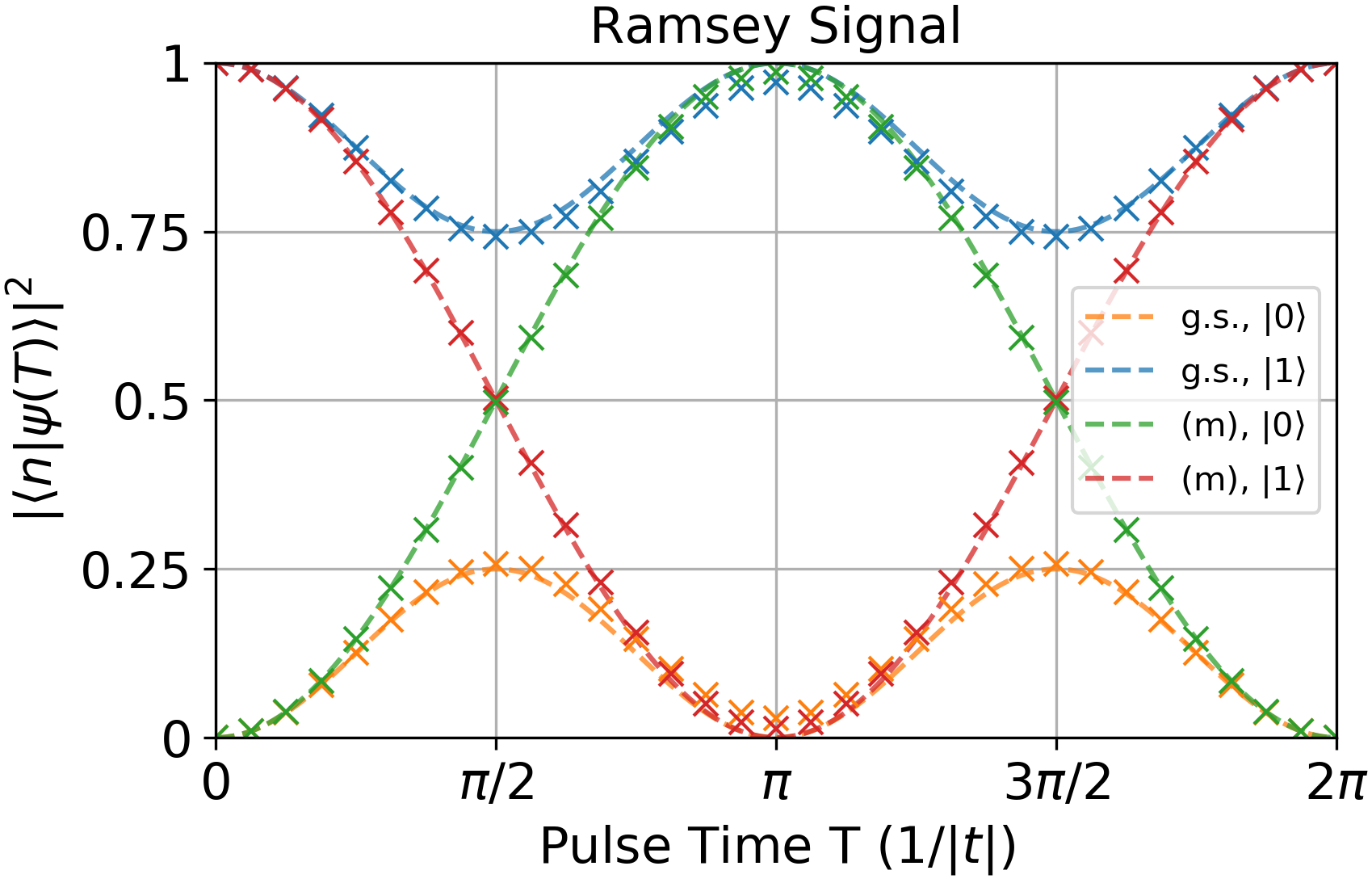, width=0.47\textwidth}
\caption{Ramsey $\pi$-pulse. Since the coupling strength of the perturbation is typically not known, the $\pi$-time has to be found experimentally. The pulse sequence is a simultaneous pulse of $\lambda_{1}$ and $\lambda_3$ followed by a simultaneous pulse of $\lambda_{2}$ and $\lambda_4$ as defined in Eq.~(\ref{eq:ramsey_operator}). The plot shows the expected curves for the measurement on the control qubit after different times $T$ for which the perturbation $\hat{V}(t)$ is applied. In the presence of a magnetic $(m)$-excitation the oscillation period doubled compared to the ground state (no excitation). The crosses show an ED simulation of the measurement protocol with a system that was prepared as discussed in Sec.~\ref{supp:growing_exc}. Therefore, also in the presence of residual excitations and imperfections in the state preparation, the $\pi$-time can be extracted robustly in the experiment.} 
\label{fig:supp_ramsey}
\end{figure}

The claim is that by braiding an electric excitation~($e$) around a magnetic excitation~($m$) the wavefunction picks up a phase $e^{i\pi}$. In order to detect this phase we suggest a Ramsey interferometric protocol~(Sec.~\ref{secTopExcBraiding}). We will calculate the expected Ramsey signal analytically and underline the robustness of the scheme by comparing it to numerical simulations using our realistically grown toric-code states, see Fig.~\ref{fig:supp_growing_mag_exc}.

As explained in the main text, a control qubit $\ket{0}$, $\ket{1}$ is introduced together with a time-dependent perturbation $\hat{V}(t)$ as defined in Eq.~(\ref{eq:ramsey_operator}). We label the states by $\ket{m=\pm 1}$ depending on the the sign of the \Ztwo~magnetic flux~$\B_P=\pm1$ of the center plaquette~(see Fig.~\ref{fig:braiding}). The time-evolution operator is given by:
\begin{align}
    \U(t,0) &= \hat{\mathcal{T}}e^{-i\int_0^t\hat{V}(\tilde{t})d\tilde{t}}\\
    \ket{\psi(t)}_I &= \U(t,0)\ket{\psi(0)}_I,
\end{align}
where $\ket{\cdot}_I$ describes a state in the interaction picture.
The time-evolution then can be computed as
\begin{align}
    \begin{split}
    \U(t,0) \ket{\psi}_\mathrm{I} &= \frac{1}{\sqrt{2}}\cos^2{(t/2)}\Big[\ket{0}+\ket{1}\Big]\otimes\ket{m=\pm1}_\mathrm{I}\\
    &-\frac{1}{\sqrt{2}}\sin^2{(t/2)}\Big[\ket{0}\pm\ket{1}\Big]\otimes\ket{m=\pm1}_\mathrm{I}\\
    &+\frac{i}{\sqrt{2}}\sin{(t/2)}\cos{(t/2)}\Big[\ket{0}\otimes(\hat{\tau}^z_1+\hat{\tau}^z_2)\\
    &+\ket{1}\otimes(\hat{\tau}^z_3+\hat{\tau}^z_4)\Big]\ket{m=\pm1}_\mathrm{I}.
    \end{split}
\end{align}

Since the measurement will be a projective measurement on the control qubit, we trace-out the bulk system from the density matrix:
    \begin{align}
    &\hat{\rho}_\mathrm{ramsey}(t) = \tr_\mathrm{bulk}\Big( \Ud(t,0) \ket{\psi}_\mathrm{I} \bra{\psi} \U(t,0)\Big)
    \\&=\frac{1}{2}\begin{pmatrix}
    1 & \cos^4{(t/2)} \pm \sin^4{(t/2)}\\
    \cos^4{(t/2)} \pm \sin^4{(t/2)} & 1
    \end{pmatrix}\nonumber
\end{align}

From $\hat{\rho}_\mathrm{ramsey}(t)$ we can conclude two things. On the one hand, we predict occupations of~$\ket{0}$ and~$\ket{1}$ versus~$t$ which is interesting because in an experiment the $\pi$-time can fluctuate and is a priori not known (see Fig.~\ref{fig:supp_ramsey}).

On the other hand, assuming a perfect~$\pi$-pulse, we can predicted the Ramsey curve from the main text (Fig.~\ref{fig:braiding}):
\begin{align}
    \hat{\rho}_\mathrm{rot}(\varphi) &= \hat{R}^\dagger_z(\varphi)\hat{\rho}_\mathrm{ramsey}(t=\pi)\hat{R}_z(\varphi)\\ 
    &=\frac{1}{2}\begin{pmatrix}
    1 & \pm e^{-i\varphi}\\
    \pm e^{-i\varphi} &  1
    \end{pmatrix}\nonumber
\end{align}
\begin{align}
    \hat{\rho}_\mathrm{measure}(\varphi) &= \hat{R}^\dagger_y(\pi/2)\hat{\rho}_\mathrm{rot}(\varphi)\hat{R}_y(\pi/2)\\ 
    &=\pm\frac{1}{2}\begin{pmatrix}
     \pm1- \cos{(\varphi)}& -i\sin{(\varphi)}\\
    i\sin{(\varphi)} &  \pm1+\cos{(\varphi)}
    \end{pmatrix}\nonumber,
\end{align}
where the operators $\hat{R}_{z,y}(\varphi)$ rotate the Bloch vector around the $z$-axis ($y$-axis) by an angle $\varphi$ ($\pi/2$). The matrix element $\bra{1}\hat{\rho}_\mathrm{measure}(\varphi)\ket{1}= \frac{1}{2}(1\pm\cos{(\varphi)})$ is plotted in Fig.~\ref{fig:braiding}.

%

\end{document}

%% file: Table_2nd_order_new.tex
\begin{tabularx}{\linewidth}{|c|c|c|c|c|c|c|X|}\specialrule{.2em}{.1em}{.1em} 

\multicolumn{1}{|c|}{\multirow{3}{*}{}}
&\multicolumn{1}{c|}{\multirow{3}{*}{initial state}}
& \multicolumn{1}{c|}{virtual state}
& \multicolumn{1}{c|}{virtual state}
& \multicolumn{1}{c|}{virtual state}
& \multicolumn{1}{c|}{virtual state}
& \multicolumn{1}{c|}{\multirow{3}{*}{final state}}
&\multicolumn{1}{c|}{\multirow{3}{*}{$\H_\mathrm{eff}$}}
\\\cline{3-6}

&
& \multicolumn{1}{c|}{coupling}
& \multicolumn{1}{c|}{coupling}
& \multicolumn{1}{c|}{coupling}
& \multicolumn{1}{c|}{coupling}
& 
& 
\\\cline{3-6}

&
& \multicolumn{1}{c|}{$\begin{aligned} ~\\[-4ex] &\Delta E \\[-2ex] \end{aligned}$}
& \multicolumn{1}{c|}{$\begin{aligned} ~\\[-4ex] &\Delta E \\[-2ex] \end{aligned}$}
& \multicolumn{1}{c|}{$\begin{aligned} ~\\[-4ex] &\Delta E \\[-2ex] \end{aligned}$}
& \multicolumn{1}{c|}{$\begin{aligned} ~\\[-4ex] &\Delta E \\[-2ex] \end{aligned}$}
&
& 
\\

\specialrule{.2em}{.1em}{.1em} 
\multicolumn{1}{|c|}{\multirow{3}{*}{$\begin{aligned}[t]~\\1\end{aligned}$}}
&\multicolumn{1}{c|}{\multirow{3}{*}{$\begin{aligned}[t]~\\ &\ket{\satisfaction{1}~_{\satisfaction{1}}^{\satisfaction{0}}~\satisfaction{0}} \end{aligned}$}} 
& \multicolumn{1}{c|}{$\begin{aligned}[t]~\\[-3.5ex] \ket{\satisfaction{0}~_{\satisfaction{4}}^{\satisfaction{0}}~\satisfaction{0}} \\[-2ex] \end{aligned}$}
& \multicolumn{1}{c|}{$\begin{aligned}[t]~\\[-3.5ex] \ket{\satisfaction{1}~_{\satisfaction{0}}^{\satisfaction{0}}~\satisfaction{1}} \\[-2ex] \end{aligned}$}
& \multicolumn{1}{c|}{$\begin{aligned}[t]~\\[-3.5ex] \ket{\satisfaction{0}~_{\satisfaction{1}}^{\satisfaction{1}}~\satisfaction{0}} \\[-2ex] \end{aligned}$}
& \multicolumn{1}{c|}{ - }
& \multicolumn{1}{c|}{\multirow{3}{*}{$\begin{aligned}[t]~\\ &\ket{\satisfaction{0}~_{\satisfaction{1}}^{\satisfaction{0}}~\satisfaction{1}} \end{aligned}$}} 
& \multicolumn{1}{c|}{\multirow{3}{*}{$\begin{aligned}[t] ~\\[-3ex]&\Big(-\frac{2g^2}{\Delta-\beta} + \frac{2g^2}{\Delta}\Big) \times\\
& \big(\ad \b +\mathrm{H.c}\big)\otimes\frac{1}{2}\big(\hat{\tau}^z + \mathbb{1}\big)\end{aligned}$ }}
\\[5pt]\cline{3-6}

&
& \multicolumn{1}{c|}{$\begin{aligned} ~\\[-4ex] &2g^2 \\[-2ex] \end{aligned}$ }
& \multicolumn{1}{c|}{$\begin{aligned} ~\\[-4ex] &g^2 \\[-2ex] \end{aligned}$ }
& \multicolumn{1}{c|}{$\begin{aligned} ~\\[-4ex] &g^2 \\[-2ex] \end{aligned}$ }
& \multicolumn{1}{c|}{ - }
&
&  \\[5pt]\cline{3-6}

&
& \multicolumn{1}{c|}{$\begin{aligned} ~\\[-4ex] &-\Delta+\beta \\[-2ex] \end{aligned}$ }
& \multicolumn{1}{c|}{$\begin{aligned} ~\\[-4ex] &\Delta \\[-2ex] \end{aligned}$ }
& \multicolumn{1}{c|}{$\begin{aligned} ~\\[-4ex] &\Delta \\[-2ex] \end{aligned}$ }
& \multicolumn{1}{c|}{ - }
&
&  \\[5pt]
\specialrule{.2em}{.1em}{.1em} 

\multicolumn{1}{|c|}{\multirow{3}{*}{$\begin{aligned}[t]~\\2\end{aligned}$}}
&\multicolumn{1}{c|}{\multirow{3}{*}{$\begin{aligned}[t]~\\ &\ket{\satisfaction{1}~_{\satisfaction{0}}^{\satisfaction{1}}~\satisfaction{0}} \end{aligned}$}} 
& \multicolumn{1}{c|}{$\begin{aligned}[t]~\\[-3.5ex] \ket{\satisfaction{0}~_{\satisfaction{0}}^{\satisfaction{4}}~\satisfaction{0}} \\[-2ex] \end{aligned}$}
& \multicolumn{1}{c|}{$\begin{aligned}[t]~\\[-3.5ex] \ket{\satisfaction{1}~_{\satisfaction{0}}^{\satisfaction{0}}~\satisfaction{1}} \\[-2ex] \end{aligned}$}
& \multicolumn{1}{c|}{$\begin{aligned}[t]~\\[-3.5ex] \ket{\satisfaction{0}~_{\satisfaction{1}}^{\satisfaction{1}}~\satisfaction{0}} \\[-2ex] \end{aligned}$}
& \multicolumn{1}{c|}{ - }
& \multicolumn{1}{c|}{\multirow{3}{*}{$\begin{aligned}[t]~\\ &\ket{\satisfaction{0}~_{\satisfaction{0}}^{\satisfaction{1}}~\satisfaction{1}} \end{aligned}$}} 
& \multicolumn{1}{c|}{\multirow{3}{*}{$\begin{aligned}[t] ~\\[-3ex]&\Big(\frac{2g^2}{\Delta-\beta} - \frac{2g^2}{\Delta}\Big) \times\\
& \big(\ad \b +\mathrm{H.c}\big)\otimes\frac{1}{2}\big( \mathbb{1} - \hat{\tau}^z \big)\end{aligned}$}}
\\[5pt]\cline{3-6}

&
& \multicolumn{1}{c|}{$\begin{aligned} ~\\[-4ex] &-2g^2 \\[-2ex] \end{aligned}$ }
& \multicolumn{1}{c|}{$\begin{aligned} ~\\[-4ex] &-g^2 \\[-2ex] \end{aligned}$ }
& \multicolumn{1}{c|}{$\begin{aligned} ~\\[-4ex] &-g^2 \\[-2ex] \end{aligned}$ }
& \multicolumn{1}{c|}{ - }
&
&  \\[5pt]\cline{3-6}

&
& \multicolumn{1}{c|}{$\begin{aligned} ~\\[-4ex] &-\Delta+\beta \\[-2ex] \end{aligned}$ }
& \multicolumn{1}{c|}{$\begin{aligned} ~\\[-4ex] &\Delta \\[-2ex] \end{aligned}$ }
& \multicolumn{1}{c|}{$\begin{aligned} ~\\[-4ex] &\Delta \\[-2ex] \end{aligned}$ }
& \multicolumn{1}{c|}{ - }
&
&  \\[5pt]
\specialrule{.2em}{.1em}{.1em} 

\multicolumn{1}{|c|}{\multirow{3}{*}{$\begin{aligned}[t]~\\3\end{aligned}$}}
&\multicolumn{1}{c|}{\multirow{3}{*}{$\begin{aligned}[t]~\\ &\ket{\satisfaction{1}~_{\satisfaction{1}}^{\satisfaction{0}}~\satisfaction{0}} \end{aligned}$}} 
& \multicolumn{1}{c|}{$\begin{aligned}[t]~\\[-3.5ex] \ket{\satisfaction{1}~_{\satisfaction{0}}^{\satisfaction{0}}~\satisfaction{1}} \\[-2ex] \end{aligned}$}
& \multicolumn{1}{c|}{$\begin{aligned}[t]~\\[-3.5ex]\ket{\satisfaction{0}~_{\satisfaction{1}}^{\satisfaction{1}}~\satisfaction{0}} \\[-2ex] \end{aligned}$}
&\multicolumn{1}{c|}{-}
& \multicolumn{1}{c|}{-}
& \multicolumn{1}{c|}{\multirow{3}{*}{$\begin{aligned}[t]~\\& \ket{\satisfaction{0}~_{\satisfaction{0}}^{\satisfaction{1}}~\satisfaction{1}} \end{aligned}$}} 
& \multicolumn{1}{c|}{\multirow{3}{*}{$\begin{aligned}[t] ~\\ &0 \end{aligned}$}}
\\[5pt]\cline{3-6}

&
& \multicolumn{1}{c|}{$\begin{aligned} ~\\[-4ex] &g^2 \\[-2ex] \end{aligned}$ }
& \multicolumn{1}{c|}{$\begin{aligned} ~\\[-4ex] &g^2 \\[-2ex] \end{aligned}$ }
& \multicolumn{1}{c|}{ - }
& \multicolumn{1}{c|}{ - }
&
&  \\[5pt]\cline{3-6}

&
& \multicolumn{1}{c|}{$\begin{aligned} ~\\[-4ex] &\Delta \\[-2ex] \end{aligned}$ }
& \multicolumn{1}{c|}{$\begin{aligned} ~\\[-4ex] &-\Delta \\[-2ex] \end{aligned}$ }
& \multicolumn{1}{c|}{ - }
& \multicolumn{1}{c|}{ - }
&
&  \\[5pt]
\specialrule{.2em}{.1em}{.1em} 

\multicolumn{1}{|c|}{\multirow{3}{*}{$\begin{aligned}[t]~\\4\end{aligned}$}}
&\multicolumn{1}{c|}{\multirow{3}{*}{$\begin{aligned}[t]~\\ &\ket{\satisfaction{1}~_{\satisfaction{0}}^{\satisfaction{1}}~\satisfaction{0}} \end{aligned}$}} 
& \multicolumn{1}{c|}{$\begin{aligned}[t]~\\[-3.5ex] \ket{\satisfaction{1}~_{\satisfaction{0}}^{\satisfaction{0}}~\satisfaction{1}} \\[-2ex] \end{aligned}$}
& \multicolumn{1}{c|}{$\begin{aligned}[t]~\\[-3.5ex]\ket{\satisfaction{0}~_{\satisfaction{1}}^{\satisfaction{1}}~\satisfaction{0}} \\[-2ex] \end{aligned}$}
&\multicolumn{1}{c|}{-}
& \multicolumn{1}{c|}{-}
& \multicolumn{1}{c|}{\multirow{3}{*}{$\begin{aligned}[t]~\\& \ket{\satisfaction{0}~_{\satisfaction{1}}^{\satisfaction{0}}~\satisfaction{1}} \end{aligned}$}} 
& \multicolumn{1}{c|}{\multirow{3}{*}{$\begin{aligned}[t] ~\\ &0 \end{aligned}$}}
\\[5pt]\cline{3-6}

&
& \multicolumn{1}{c|}{$\begin{aligned} ~\\[-4ex] &-g^2 \\[-2ex] \end{aligned}$ }
& \multicolumn{1}{c|}{$\begin{aligned} ~\\[-4ex] &-g^2 \\[-2ex] \end{aligned}$ }
& \multicolumn{1}{c|}{ - }
& \multicolumn{1}{c|}{ - }
&
&  \\[5pt]\cline{3-6}

&
& \multicolumn{1}{c|}{$\begin{aligned} ~\\[-4ex] &\Delta \\[-2ex] \end{aligned}$ }
& \multicolumn{1}{c|}{$\begin{aligned} ~\\[-4ex] &-\Delta \\[-2ex] \end{aligned}$ }
& \multicolumn{1}{c|}{ - }
& \multicolumn{1}{c|}{ - }
&
&  \\[5pt]
\specialrule{.2em}{.1em}{.1em} 

\multicolumn{1}{|c|}{\multirow{3}{*}{$\begin{aligned}[t]~\\5\end{aligned}$}}
&\multicolumn{1}{c|}{\multirow{3}{*}{$\begin{aligned}[t]~\\ &\ket{\satisfaction{1}~_{\satisfaction{1}}^{\satisfaction{0}}~\satisfaction{0}} \end{aligned}$}} 
& \multicolumn{1}{c|}{$\begin{aligned}[t]~\\[-3.5ex] \ket{\satisfaction{4}~_{\satisfaction{0}}^{\satisfaction{0}}~\satisfaction{0}} \\[-2ex] \end{aligned}$}
& \multicolumn{1}{c|}{$\begin{aligned}[t]~\\[-3.5ex]\ket{\satisfaction{0}~_{\satisfaction{1}}^{\satisfaction{1}}~\satisfaction{0}} \\[-2ex] \end{aligned}$}
&\multicolumn{1}{c|}{$\begin{aligned}[t]~\\[-3.5ex]\ket{\satisfaction{1}~_{\satisfaction{0}}^{\satisfaction{0}}~\satisfaction{1}} \\[-2ex] \end{aligned}$}
& \multicolumn{1}{c|}{-}
& \multicolumn{1}{c|}{\multirow{3}{*}{$\begin{aligned}[t]~\\& \ket{\satisfaction{1}~_{\satisfaction{0}}^{\satisfaction{1}}~\satisfaction{0}} \end{aligned}$}} 
& \multicolumn{1}{c|}{\multirow{3}{*}{$\begin{aligned}[t] ~\\[-3ex]&\Big(\frac{2g^2}{\Delta+\alpha} - \frac{2g^2}{\Delta}\Big) \times\\
& \ad \a \big(1 - \bd \b\big)\otimes\hat{\tau}^x \end{aligned}$ }}
\\[5pt]\cline{3-6}

&
& \multicolumn{1}{c|}{$\begin{aligned} ~\\[-4ex] &2g^2 \\[-2ex] \end{aligned}$ }
& \multicolumn{1}{c|}{$\begin{aligned} ~\\[-4ex] &g^2 \\[-2ex] \end{aligned}$ }
& \multicolumn{1}{c|}{$\begin{aligned} ~\\[-4ex] &-g^2 \\[-2ex] \end{aligned}$ }
& \multicolumn{1}{c|}{ - }
&
&  \\[5pt]\cline{3-6}

&
& \multicolumn{1}{c|}{$\begin{aligned} ~\\[-4ex] &\Delta+\alpha \\[-2ex] \end{aligned}$ }
& \multicolumn{1}{c|}{$\begin{aligned} ~\\[-4ex] &-\Delta \\[-2ex] \end{aligned}$ }
& \multicolumn{1}{c|}{$\begin{aligned} ~\\[-4ex] &\Delta \\[-2ex] \end{aligned}$ }
& \multicolumn{1}{c|}{ - }
&
&  \\[5pt]
\specialrule{.2em}{.1em}{.1em} 

\multicolumn{1}{|c|}{\multirow{3}{*}{$\begin{aligned}[t]~\\6\end{aligned}$}}
&\multicolumn{1}{c|}{\multirow{3}{*}{$\begin{aligned}[t]~\\ &\ket{\satisfaction{0}~_{\satisfaction{1}}^{\satisfaction{0}}~\satisfaction{1}} \end{aligned}$}} 
& \multicolumn{1}{c|}{$\begin{aligned}[t]~\\[-3.5ex] \ket{\satisfaction{0}~_{\satisfaction{0}}^{\satisfaction{0}}~\satisfaction{4}} \\[-2ex] \end{aligned}$}
& \multicolumn{1}{c|}{$\begin{aligned}[t]~\\[-3.5ex]\ket{\satisfaction{0}~_{\satisfaction{1}}^{\satisfaction{1}}~\satisfaction{0}} \\[-2ex] \end{aligned}$}
&\multicolumn{1}{c|}{$\begin{aligned}[t]~\\[-3.5ex]\ket{\satisfaction{1}~_{\satisfaction{0}}^{\satisfaction{0}}~\satisfaction{1}} \\[-2ex] \end{aligned}$}
& \multicolumn{1}{c|}{-}
& \multicolumn{1}{c|}{\multirow{3}{*}{$\begin{aligned}[t]~\\& \ket{\satisfaction{0}~_{\satisfaction{0}}^{\satisfaction{1}}~\satisfaction{1}} \end{aligned}$}} 
& \multicolumn{1}{c|}{\multirow{3}{*}{$\begin{aligned}[t] ~\\[-3ex]&\Big(\frac{2g^2}{\Delta+\alpha} - \frac{2g^2}{\Delta}\Big) \times\\
& \big(1 - \ad \a\big)\bd \b \otimes\hat{\tau}^x \end{aligned}$ }}
\\[5pt]\cline{3-6}

&
& \multicolumn{1}{c|}{$\begin{aligned} ~\\[-4ex] &-2g^2 \\[-2ex] \end{aligned}$ }
& \multicolumn{1}{c|}{$\begin{aligned} ~\\[-4ex] &-g^2 \\[-2ex] \end{aligned}$ }
& \multicolumn{1}{c|}{$\begin{aligned} ~\\[-4ex] &g^2 \\[-2ex] \end{aligned}$ }
& \multicolumn{1}{c|}{ - }
&
&  \\[5pt]\cline{3-6}

&
& \multicolumn{1}{c|}{$\begin{aligned} ~\\[-4ex] &\Delta+\alpha \\[-2ex] \end{aligned}$ }
& \multicolumn{1}{c|}{$\begin{aligned} ~\\[-4ex] &-\Delta \\[-2ex] \end{aligned}$ }
& \multicolumn{1}{c|}{$\begin{aligned} ~\\[-4ex] &\Delta \\[-2ex] \end{aligned}$ }
& \multicolumn{1}{c|}{ - }
&
&  \\[5pt]
\specialrule{.2em}{.1em}{.1em} 

\multicolumn{1}{|c|}{\multirow{3}{*}{$\begin{aligned}[t]~\\7\end{aligned}$}}
&\multicolumn{1}{c|}{\multirow{3}{*}{$\begin{aligned}[t]~\\[-1.5ex] &\ket{\satisfaction{1}~_{\satisfaction{1}}^{\satisfaction{0}}~\satisfaction{0}},\\
&\ket{\satisfaction{0}~_{\satisfaction{1}}^{\satisfaction{0}}~\satisfaction{1}}
\end{aligned}$}}
& \multicolumn{1}{c|}{$\begin{aligned}[t]~\\[-3.5ex] \ket{\satisfaction{4}~_{\satisfaction{0}}^{\satisfaction{0}}~\satisfaction{0}},\ket{\satisfaction{0}~_{\satisfaction{0}}^{\satisfaction{0}}~\satisfaction{4}} \\[-2ex] \end{aligned}$}
& \multicolumn{1}{c|}{$\begin{aligned}[t]~\\[-3.5ex]\ket{\satisfaction{0}~_{\satisfaction{4}}^{\satisfaction{0}}~\satisfaction{0}} \\[-2ex] \end{aligned}$}
&\multicolumn{1}{c|}{$\begin{aligned}[t]~\\[-3.5ex]\ket{\satisfaction{0}~_{\satisfaction{1}}^{\satisfaction{1}}~\satisfaction{0}} \\[-2ex] \end{aligned}$}
& \multicolumn{1}{c|}{$\begin{aligned}[t]~\\[-3.5ex]\ket{\satisfaction{1}~_{\satisfaction{0}}^{\satisfaction{0}}~\satisfaction{1}} \\[-2ex] \end{aligned}$}
& \multicolumn{1}{c|}{\multirow{3}{*}{$\begin{aligned}[t]~\\[-1.5ex] &\ket{\satisfaction{1}~_{\satisfaction{1}}^{\satisfaction{0}}~\satisfaction{0}},\\
&\ket{\satisfaction{0}~_{\satisfaction{1}}^{\satisfaction{0}}~\satisfaction{1}}
\end{aligned}$}} 
& \multicolumn{1}{c|}{\multirow{3}{*}{$\begin{aligned}[t] ~\\[-3ex]&\Big(\frac{2g^2}{\Delta+\alpha} - \frac{2g^2}{\Delta-\beta}\Big) \times\\
& \big(\ad \a + \bd \b\big)\otimes\frac{1}{2}\big(\hat{\tau}^z + \mathbb{1}\big) \end{aligned}$ }}
\\[5pt]\cline{3-6}

&
& \multicolumn{1}{c|}{$\begin{aligned} ~\\[-4ex] &2g^2 \\[-2ex] \end{aligned}$ }
& \multicolumn{1}{c|}{$\begin{aligned} ~\\[-4ex] &2g^2 \\[-2ex] \end{aligned}$ }
& \multicolumn{1}{c|}{$\begin{aligned} ~\\[-4ex] &g^2 \\[-2ex] \end{aligned}$ }
& \multicolumn{1}{c|}{$\begin{aligned} ~\\[-4ex] &g^2 \\[-2ex] \end{aligned}$}
&
&  \\[5pt]\cline{3-6}

&
& \multicolumn{1}{c|}{$\begin{aligned} ~\\[-4ex] &\Delta+\alpha \\[-2ex] \end{aligned}$ }
& \multicolumn{1}{c|}{$\begin{aligned} ~\\[-4ex] &-\Delta+\beta \\[-2ex] \end{aligned}$ }
& \multicolumn{1}{c|}{$\begin{aligned} ~\\[-4ex] &-\Delta \\[-2ex] \end{aligned}$ }
& \multicolumn{1}{c|}{$\begin{aligned} ~\\[-4ex] &\Delta \\[-2ex] \end{aligned}$ }
&
&  \\[5pt]
\specialrule{.2em}{.1em}{.1em} 

\multicolumn{1}{|c|}{\multirow{3}{*}{$\begin{aligned}[t]~\\8\end{aligned}$}}
&\multicolumn{1}{c|}{\multirow{3}{*}{$\begin{aligned}[t]~\\[-1.5ex] &\ket{\satisfaction{1}~_{\satisfaction{0}}^{\satisfaction{1}}~\satisfaction{0}},\\
&\ket{\satisfaction{0}~_{\satisfaction{0}}^{\satisfaction{1}}~\satisfaction{1}}
\end{aligned}$}}
& \multicolumn{1}{c|}{$\begin{aligned}[t]~\\[-3.5ex] \ket{\satisfaction{4}~_{\satisfaction{0}}^{\satisfaction{0}}~\satisfaction{0}},\ket{\satisfaction{0}~_{\satisfaction{0}}^{\satisfaction{0}}~\satisfaction{4}} \\[-2ex] \end{aligned}$}
& \multicolumn{1}{c|}{$\begin{aligned}[t]~\\[-3.5ex]\ket{\satisfaction{0}~_{\satisfaction{0}}^{\satisfaction{4}}~\satisfaction{0}} \\[-2ex] \end{aligned}$}
&\multicolumn{1}{c|}{$\begin{aligned}[t]~\\[-3.5ex]\ket{\satisfaction{0}~_{\satisfaction{1}}^{\satisfaction{1}}~\satisfaction{0}} \\[-2ex] \end{aligned}$}
& \multicolumn{1}{c|}{$\begin{aligned}[t]~\\[-3.5ex]\ket{\satisfaction{1}~_{\satisfaction{0}}^{\satisfaction{0}}~\satisfaction{1}} \\[-2ex] \end{aligned}$}
& \multicolumn{1}{c|}{\multirow{3}{*}{$\begin{aligned}[t]~\\[-1.5ex] &\ket{\satisfaction{1}~_{\satisfaction{0}}^{\satisfaction{1}}~\satisfaction{0}},\\
&\ket{\satisfaction{0}~_{\satisfaction{0}}^{\satisfaction{1}}~\satisfaction{1}}
\end{aligned}$}} 
& \multicolumn{1}{c|}{\multirow{3}{*}{$\begin{aligned}[t] ~\\[-3ex]&\Big(\frac{2g^2}{\Delta+\alpha} - \frac{2g^2}{\Delta-\beta}\Big) \times\\
& \big(\ad \a + \bd \b\big)\otimes\frac{1}{2}\big(\mathbb{1} - \hat{\tau}^z\big) \end{aligned}$ }}
\\[5pt]\cline{3-6}

&
& \multicolumn{1}{c|}{$\begin{aligned} ~\\[-4ex] &2g^2 \\[-2ex] \end{aligned}$ }
& \multicolumn{1}{c|}{$\begin{aligned} ~\\[-4ex] &2g^2 \\[-2ex] \end{aligned}$ }
& \multicolumn{1}{c|}{$\begin{aligned} ~\\[-4ex] &g^2 \\[-2ex] \end{aligned}$ }
& \multicolumn{1}{c|}{$\begin{aligned} ~\\[-4ex] &g^2 \\[-2ex] \end{aligned}$}
&
&  \\[5pt]\cline{3-6}

&
& \multicolumn{1}{c|}{$\begin{aligned} ~\\[-4ex] &\Delta+\alpha \\[-2ex] \end{aligned}$ }
& \multicolumn{1}{c|}{$\begin{aligned} ~\\[-4ex] &-\Delta+\beta \\[-2ex] \end{aligned}$ }
& \multicolumn{1}{c|}{$\begin{aligned} ~\\[-4ex] &-\Delta \\[-2ex] \end{aligned}$ }
& \multicolumn{1}{c|}{$\begin{aligned} ~\\[-4ex] &\Delta \\[-2ex] \end{aligned}$ }
&
&  \\[5pt]
\specialrule{.35em}{.1em}{.1em} 

\multicolumn{1}{|c|}{\multirow{3}{*}{$\begin{aligned}[t]~\\9\end{aligned}$}}
&\multicolumn{1}{c|}{\multirow{3}{*}{$\begin{aligned}[t]~\\ &\ket{\satisfaction{1}~_{\satisfaction{1}}^{\satisfaction{0}}~\satisfaction{1}} \end{aligned}$}} 
& \multicolumn{1}{c|}{$\begin{aligned}[t]~\\[-3.5ex] \ket{\satisfaction{4}~_{\satisfaction{0}}^{\satisfaction{0}}~\satisfaction{1}} \\[-2ex] \end{aligned}$}
& \multicolumn{1}{c|}{$\begin{aligned}[t]~\\[-3.5ex]\ket{\satisfaction{0}~_{\satisfaction{1}}^{\satisfaction{1}}~\satisfaction{1}} \\[-2ex] \end{aligned}$}
&\multicolumn{1}{c|}{$\begin{aligned}[t]~\\[-3.5ex]\ket{\satisfaction{1}~_{\satisfaction{0}}^{\satisfaction{0}}~\satisfaction{4}} \\[-2ex] \end{aligned}$}
& \multicolumn{1}{c|}{$\begin{aligned}[t]~\\[-3.5ex]\ket{\satisfaction{1}~_{\satisfaction{1}}^{\satisfaction{1}}~\satisfaction{0}} \\[-2ex] \end{aligned}$}
& \multicolumn{1}{c|}{\multirow{3}{*}{$\begin{aligned}[t]~\\& \ket{\satisfaction{1}~_{\satisfaction{0}}^{\satisfaction{1}}~\satisfaction{1}} \end{aligned}$}} 
& \multicolumn{1}{c|}{\multirow{3}{*}{$\begin{aligned}[t] ~\\ &0 \end{aligned}$ }}
\\[5pt]\cline{3-6}

&
& \multicolumn{1}{c|}{$\begin{aligned} ~\\[-4ex] &2g^2 \\[-2ex] \end{aligned}$ }
& \multicolumn{1}{c|}{$\begin{aligned} ~\\[-4ex] &g^2 \\[-2ex] \end{aligned}$ }
& \multicolumn{1}{c|}{$\begin{aligned} ~\\[-4ex] &-2g^2 \\[-2ex] \end{aligned}$ }
& \multicolumn{1}{c|}{$\begin{aligned} ~\\[-4ex] &-g^2 \\[-2ex] \end{aligned}$ }
&
&  \\[5pt]\cline{3-6}

&
& \multicolumn{1}{c|}{$\begin{aligned} ~\\[-4ex] &\Delta+\alpha \\[-2ex] \end{aligned}$ }
& \multicolumn{1}{c|}{$\begin{aligned} ~\\[-4ex] &-\Delta \\[-2ex] \end{aligned}$ }
& \multicolumn{1}{c|}{$\begin{aligned} ~\\[-4ex] &\Delta+\alpha \\[-2ex] \end{aligned}$ }
& \multicolumn{1}{c|}{$\begin{aligned} ~\\[-4ex] &-\Delta \\[-2ex] \end{aligned}$ }
&
&  \\[5pt]
\specialrule{.2em}{.1em}{.1em} 

\multicolumn{1}{|c|}{\multirow{3}{*}{$\begin{aligned}[t]~\\10\end{aligned}$}}
&\multicolumn{1}{c|}{\multirow{3}{*}{$\begin{aligned}[t]~\\[-1.5ex] &\ket{\satisfaction{1}~_{\satisfaction{1}}^{\satisfaction{0}}~\satisfaction{1}},\\
&\ket{\satisfaction{1}~_{\satisfaction{0}}^{\satisfaction{1}}~\satisfaction{1}}\end{aligned}$}} 
& \multicolumn{1}{c|}{$\begin{aligned}[t]~\\[-3.5ex] \ket{\satisfaction{4}~_{\satisfaction{0}}^{\satisfaction{0}}~\satisfaction{1}} \\[-2ex] \end{aligned}$}
& \multicolumn{1}{c|}{$\begin{aligned}[t]~\\[-3.5ex]\ket{\satisfaction{0}~_{\satisfaction{4}}^{\satisfaction{0}}~\satisfaction{1}},\ket{\satisfaction{0}~_{\satisfaction{0}}^{\satisfaction{4}}~\satisfaction{1}} \\[-2ex] \end{aligned}$}
&\multicolumn{1}{c|}{$\begin{aligned}[t]~\\[-3.5ex]\ket{\satisfaction{1}~_{\satisfaction{0}}^{\satisfaction{0}}~\satisfaction{4}} \\[-2ex] \end{aligned}$}
& \multicolumn{1}{c|}{$\begin{aligned}[t]~\\[-3.5ex]\ket{\satisfaction{1}~_{\satisfaction{4}}^{\satisfaction{0}}~\satisfaction{0}},\ket{\satisfaction{1}~_{\satisfaction{0}}^{\satisfaction{4}}~\satisfaction{0}} \\[-2ex] \end{aligned}$}
& \multicolumn{1}{c|}{\multirow{3}{*}{$\begin{aligned}[t]~\\[-1.5ex] &\ket{\satisfaction{1}~_{\satisfaction{1}}^{\satisfaction{0}}~\satisfaction{1}},\\
&\ket{\satisfaction{1}~_{\satisfaction{0}}^{\satisfaction{1}}~\satisfaction{1}}\end{aligned}$}} 
& \multicolumn{1}{c|}{\multirow{3}{*}{$\begin{aligned}[t] ~\\[-3ex]\Big(\frac{2g^2}{\Delta+\alpha} &- \frac{2g^2}{\Delta-\beta}\Big) \times\\
 &\ad \a \bd \b \otimes\mathbb{1} \end{aligned}$ }}
\\[5pt]\cline{3-6}

&
& \multicolumn{1}{c|}{$\begin{aligned} ~\\[-4ex] &2g^2 \\[-2ex] \end{aligned}$ }
& \multicolumn{1}{c|}{$\begin{aligned} ~\\[-4ex] &2g^2 \\[-2ex] \end{aligned}$ }
& \multicolumn{1}{c|}{$\begin{aligned} ~\\[-4ex] &2g^2 \\[-2ex] \end{aligned}$ }
& \multicolumn{1}{c|}{$\begin{aligned} ~\\[-4ex] &2g^2 \\[-2ex] \end{aligned}$ }
&
&  \\[5pt]\cline{3-6}

&
& \multicolumn{1}{c|}{$\begin{aligned} ~\\[-4ex] &\Delta+\alpha \\[-2ex] \end{aligned}$ }
& \multicolumn{1}{c|}{$\begin{aligned} ~\\[-4ex] &-\Delta+\beta \\[-2ex] \end{aligned}$ }
& \multicolumn{1}{c|}{$\begin{aligned} ~\\[-4ex] &\Delta+\alpha \\[-2ex] \end{aligned}$ }
& \multicolumn{1}{c|}{$\begin{aligned} ~\\[-4ex] &-\Delta+\beta \\[-2ex] \end{aligned}$ }
&
&  \\[5pt]
\specialrule{.35em}{.1em}{.1em} 

\multicolumn{1}{|c|}{\multirow{3}{*}{$\begin{aligned}[t]~\\11\end{aligned}$}}
&\multicolumn{1}{c|}{\multirow{3}{*}{$\begin{aligned}[t]~\\[-1.5ex] &\ket{\satisfaction{0}~_{\satisfaction{1}}^{\satisfaction{0}}~\satisfaction{0}},\\
&\ket{\satisfaction{0}~_{\satisfaction{0}}^{\satisfaction{1}}~\satisfaction{0}}
\end{aligned}$}}
& \multicolumn{1}{c|}{$\begin{aligned}[t]~\\[-3.5ex] \ket{\satisfaction{1}~_{\satisfaction{0}}^{\satisfaction{0}}~\satisfaction{0}} \\[-2ex] \end{aligned}$}
& \multicolumn{1}{c|}{$\begin{aligned}[t]~\\[-3.5ex] \ket{\satisfaction{0}~_{\satisfaction{0}}^{\satisfaction{0}}~\satisfaction{1}} \\[-2ex] \end{aligned}$}
&\multicolumn{1}{c|}{ - }
& \multicolumn{1}{c|}{ - }
& \multicolumn{1}{c|}{\multirow{3}{*}{$\begin{aligned}[t]~\\[-1.5ex] &\ket{\satisfaction{0}~_{\satisfaction{1}}^{\satisfaction{0}}~\satisfaction{0}},\\
&\ket{\satisfaction{0}~_{\satisfaction{0}}^{\satisfaction{1}}~\satisfaction{0}}
\end{aligned}$}} 
& \multicolumn{1}{c|}{\multirow{3}{*}{$\begin{aligned}[t] ~\\[-3ex] \frac{2g^2}{\Delta}&\times\\
 \big(1 - \ad \a \big)&\big(1 - \bd \b \big)\otimes \mathbb{1} \end{aligned}$ }}
\\[5pt]\cline{3-6}

&
& \multicolumn{1}{c|}{$\begin{aligned} ~\\[-4ex] &g^2 \\[-2ex] \end{aligned}$ }
& \multicolumn{1}{c|}{$\begin{aligned} ~\\[-4ex] &g^2 \\[-2ex] \end{aligned}$ }
& \multicolumn{1}{c|}{ - }
& \multicolumn{1}{c|}{ -}
&
&  \\[5pt]\cline{3-6}

&
& \multicolumn{1}{c|}{$\begin{aligned} ~\\[-4ex] &\Delta \\[-2ex] \end{aligned}$ }
& \multicolumn{1}{c|}{$\begin{aligned} ~\\[-4ex] &\Delta \\[-2ex] \end{aligned}$ }
& \multicolumn{1}{c|}{ - }
& \multicolumn{1}{c|}{ - }
&
&  \\[5pt]
\specialrule{.2em}{.1em}{.1em} 

\multicolumn{1}{|c|}{\multirow{3}{*}{$\begin{aligned}[t]~\\12\end{aligned}$}}
&\multicolumn{1}{c|}{\multirow{3}{*}{$\begin{aligned}[t]~\\[-1.5ex] &\ket{\satisfaction{0}~_{\satisfaction{1}}^{\satisfaction{0}}~\satisfaction{0}},\\
&\ket{\satisfaction{0}~_{\satisfaction{0}}^{\satisfaction{1}}~\satisfaction{0}}
\end{aligned}$}}
& \multicolumn{1}{c|}{$\begin{aligned}[t]~\\[-3.5ex] \ket{\satisfaction{1}~_{\satisfaction{0}}^{\satisfaction{0}}~\satisfaction{0}} \\[-2ex] \end{aligned}$}
& \multicolumn{1}{c|}{$\begin{aligned}[t]~\\[-3.5ex] \ket{\satisfaction{0}~_{\satisfaction{0}}^{\satisfaction{0}}~\satisfaction{1}} \\[-2ex] \end{aligned}$}
&\multicolumn{1}{c|}{ - }
& \multicolumn{1}{c|}{ - }
& \multicolumn{1}{c|}{\multirow{3}{*}{$\begin{aligned}[t]~\\[-1.5ex] &\ket{\satisfaction{0}~_{\satisfaction{0}}^{\satisfaction{1}}~\satisfaction{0}},\\
&\ket{\satisfaction{0}~_{\satisfaction{1}}^{\satisfaction{0}}~\satisfaction{0}}
\end{aligned}$}} 
& \multicolumn{1}{c|}{\multirow{3}{*}{$\begin{aligned}[t] ~\\ &0 \end{aligned}$ }}
\\[5pt]\cline{3-6}

&
& \multicolumn{1}{c|}{$\begin{aligned} ~\\[-4ex] &g^2 \\[-2ex] \end{aligned}$ }
& \multicolumn{1}{c|}{$\begin{aligned} ~\\[-4ex] &-g^2 \\[-2ex] \end{aligned}$ }
& \multicolumn{1}{c|}{ - }
& \multicolumn{1}{c|}{ -}
&
&  \\[5pt]\cline{3-6}

&
& \multicolumn{1}{c|}{$\begin{aligned} ~\\[-4ex] &\Delta \\[-2ex] \end{aligned}$ }
& \multicolumn{1}{c|}{$\begin{aligned} ~\\[-4ex] &\Delta \\[-2ex] \end{aligned}$ }
& \multicolumn{1}{c|}{ - }
& \multicolumn{1}{c|}{ - }
&
&  \\[5pt]
\specialrule{.2em}{.1em}{.1em} 

\end{tabularx}